\documentclass[twoside,11pt]{article}

\usepackage{jmlr2e}
\usepackage{graphicx}
\usepackage{amsmath}
\usepackage{color}
\usepackage{siunitx}
\usepackage{hyperref}
\usepackage{booktabs}
\usepackage{amsfonts}
\usepackage{wrapfig}
\usepackage{subcaption}
\usepackage{tikz-cd}
\usetikzlibrary{shapes,snakes,shapes.geometric}
\usetikzlibrary{decorations,arrows,calc,arrows.meta,fit,positioning}
\tikzset{
    -Latex,auto,node distance =1 cm and 1 cm,semithick,
    state/.style ={ellipse, draw, minimum width = 0.5 cm},
    point/.style = {circle, draw, inner sep=0.04cm,fill,node contents={}},
    bidirected/.style={Latex-Latex,dashed},
    el/.style = {inner sep=2pt, align=left, sloped}
}
\usepackage[linesnumbered,algoruled,boxed,lined,ruled]{algorithm2e}

\let\oldnl\nl
\newcommand{\nonl}{\renewcommand{\nl}{\let\nl\oldnl}}

\newtheorem{assumption}{Assumption}
\newtheorem{model}{Model}
\definecolor{darkblue}{rgb}{0.0, 0.0, 0.55}

\newcommand{\qlsearch}{{\sc \textbf{\texttt{QLatentSearch}}}}
\newcommand{\qinferg}{{\sc \textbf{\texttt{QInferGraph}}}}

\jmlrheading{106}{2022}{062425}{09/08/2022}{12/19/2022}{meila00a}{Mohammad Ali Javidian   and Vaneet Aggarwal and Zubin Jacob}

\ShortHeadings{Quantum Causality: an Entropic Approach}{Javidian and Aggarwal and Jacob}
\firstpageno{1}

\begin{document}

\title{Quantum Causal Inference in the Presence of Hidden Common Causes: an Entropic Approach}

\author{\name Mohammad Ali Javidian\thanks{Present address: Appalachian State University, Boone, NC, 28608} \email javidianma@appstate.edu \\
\name Vaneet Aggarwal \email vaneet@purdue.edu \\
\name Zubin Jacob \email zjacob@purdue.edu \\
       \addr School of Electrical and Computer Engineering\\
       Purdue University\\
       West Lafayette, IN 47907, USA
       }

\editor{}

\maketitle

\begin{abstract}
Quantum causality is an emerging field of study which has the potential to greatly advance our understanding of quantum systems. In this paper, we put forth a theoretical framework for merging quantum information science and causal inference by exploiting entropic principles. For this purpose, we leverage the tradeoff between the entropy of hidden cause and the conditional mutual information of observed variables to develop a scalable algorithmic approach for inferring causality in the presence of latent confounders (common causes) in quantum systems. As an application, we consider a system of three entangled qubits and transmit the second and third qubits over separate noisy quantum channels. In this model, we validate that the first qubit is a latent confounder and the common cause of the second and third qubits. In contrast, when two entangled qubits are prepared and one of them is sent over a noisy channel, there is no common confounder. We also demonstrate that the proposed approach outperforms the results of classical causal inference for the Tubingen database when the variables are classical by exploiting quantum dependence between variables through density matrices rather than joint probability distributions. Thus, the proposed approach unifies classical and quantum causal inference in a principled way.
\end{abstract}


\section{Introduction}\label{intro}

\paragraph{Motivation}
Causal inference lies at the heart of science \citep{Pearl09,pearl2018book}: the conclusions drawn from scientific studies almost always involve extracting causation (cause and effect relationships) from association, even if researchers often refrain from explicitly acknowledging the causal goal of research projects \citep{hernan2018c,hernan2019second}. However, causal inference from observational data is an ambitious and difficult task. Identifying cause and effect relationships from observational data is even more challenging in the presence
of \textit{hidden common causes} (\textit{latent confounders}) \citep{heckerman2019toward}. The broad impact of this phenomena has been studied in multiple domains of science such as epidemiologic studies \citep{lipsitch2010negative}, biology and medicine \citep{skelly2012assessing,meinshausen2016methods}, experiential education \citep{ewert2009creating,Kallus2018Removing}, economics and marketing \citep{varian2016causal,hunermund2019causal}, among others.

A similar concept is increasingly appreciated among quantum physicists, namely the inference of \textit{quantum common causes} \citep{wolfe2020quantifying,allen2017quantum,ried2015quantum,Chaves2014UAI,chaves2014causal,chaves2015information,hofer1999reichenbach}. It has been used to provide a satisfactory
causal explanation (i.e., non-fine-tuned)  of Bell inequality violations \citep{allen2017quantum,hofer1999reichenbach}. This also has led to a formalization of quantum causal models \citep{costa2016quantum,barrett2019quantum,chiribella2019quantum,shrapnel2019discovering}. As shown in \citep{Chaves2014UAI,chaves2014causal,chaves2015information}, \textit{in some cases}, (hidden) common causes can be distinguished from direct causation using information theoretical generalization of Bell’s inequalities and causal directed acyclic graphs (DAGs). Also, as shown in \citep{fitzsimons2015quantum,ried2015quantum}, observed quantum correlations alone are \textit{sometimes} enough to imply causation. However, the proposed approach in \citep{fitzsimons2015quantum,ried2015quantum} depends on the precise knowledge of the physical system and the measurement apparatuses \citep{gachechiladze2020quantifying}. In this paper, we propose the first tractable algorithmic approach to distinguish between a hidden common cause and direct causal influences among two observed quantum systems without any interventional data.

To show the difficulty of causal structure discovery task even in the simplest classical case, where our observation consists of only two jointly-distributed
random variables $X$ and $Y$ that are statically correlated, we recall Reichenbach's common cause principle \citep{reichenbach1991direction}: If two random variables $X$ and $Y$ are statistically dependent, then there exists a third variable $Z$ that causally affects both. As a special case, $Z$ may coincide with either $X$ or $Y$. Furthermore, this variable $Z$ makes $X$ and $Y$ conditionally independent, i.e.,    $X\perp\!\!\!\perp  Y\text{\textbar}Z$.  So, possible candidates for representing causal relationships between $X$ and $Y$ are: $X\to Y$, $X\gets Y$, and $X\gets Z\to Y$, and there is no easy way to determine which one is the right structure based on the observational data alone. The variable $Z$ in the case $X\gets Z\to Y$ is called \textit{unmeasured (latent) confounder}  or \textit{unmeasured (latent) common cause}. So, one of the fundamental questions in causality is to determine how cause-effect relationships can be inferred from statistical information, encoded as a joint probability distribution, obtained under normal, intervention-free experiments. 

\paragraph{Co-existence of Quantum Systems}
To discover the true cause-effect relationships, scientists normally perform randomized experiments
where a sample of units drawn from the population of interest is subjected to the
specified manipulation directly. In many cases, however, such a direct approach is
not possible due to expense or ethical considerations. Instead, investigators have
to rely on observational studies to infer causality. This task is even more challenging in quantum context due to quantum superpositions and entanglement relations. In this work, we are interested in quantum generalizations of causal structures in the presence of latent common causes. These structures can be shown as a directed acyclic graph (DAG), where nodes are quantum systems, and edges are quantum operations\footnote{In the context of quantum computation \citep{hogg1996quantum}, a quantum operation is called a quantum channel.}. However, the key theoretical distinction between an entirely classical causal structure and a quantum casual structure is the concept of \textit{coexisting}. Because of the impossibility of cloning, the outcomes and the quantum systems that led to them do not exist simultaneously. If a system $X$ is measured to produce $Y$, then $\rho_{XY}$ is not defined and hence neither is the entropy $S(\rho_{XY})$ \citep{weilenmann2017analysing}.  For a given causal structure, a \textit{coexisting set} of systems is one for which a joint state can be defined \citep{chaves2015information,weilenmann2017analysing, weilenmann2020analysing}.

If we pick a coexisting set of nodes (e.g., a classical system, or a set of nodes that are created at the same instance of time, i.e., they do enjoy a joint density operator), then we can investigate the identification of quantum causal structures in the presence of latent confounders. 

\paragraph{Contributions}
In this paper, we  consider causality between two coexisting quantum subsystems.  As a part of the evaluation framework, we provide a model of such a coexisting system, where two entangled qubits are used, and one of the  qubit is transmitted over a quantum channel. Similarly, three entangled qubits are used, and two of them are transmitted over two separate quantum channels. The models can be further generalized, while note that the subsystems which are being considered for quantum causality relationships have to coexist, unlike in the classical case where it is not necessary for the sub-systems to coexist.
To address this problem, we introduce a theoretical framework to merge quantum information science with causal inference using entropic principles. Classically, it has been proposed and tested that minimization of the trade-off between the entropy of the (hidden) common cause $Z$ (i.e., $H(Z)$) and the conditional mutual information of observed variables $X$ and $Y$ given $Z$ (i.e., $I(X;Y\text{\textbar}Z)$)  can be used to distinguish the \textit{latent graph} $X\gets Z\to Y$ ($Z$ is an unmeasured confounder) from the directed graphs $X\to Y$ and $X\gets Y$ based on observational data alone under certain assumptions \citep{KocaogluNEURIPS2020} (a brief review is given in Section \ref{sec:classic}). We will provide the first generalization of this approach to the quantum domain.

Even though the paper considers an approach for quantum causal inference, we also apply the proposed approach to a classical setup, where two bits are transmitted over a binary symmetric channel (to illustrate the case of no confounder), or two bits are transmitted over two separate channels (to illustrate the  case of latent confounder). 
We note that finding the optima over a quantum density matrix rather than over the probability distribution function provides larger degrees of freedom thus resulting in improved results. This example is used to select the hyperparameters for our framework, and these hyperparameters are used in the rest of the paper.  This demonstrates that the proposed approach can also be used for classical causal inference with improved results. Our main contributions are as follows:

\noindent$\bullet$   Inferring causality in the presence of latent confounders from observational data alone is one of the most important and challenging problems in statistical inference. We propose an iterative algorithm, called \qinferg, for identifying \textit{latent confounders} in Section \ref{sec:method}. Our method leverages the concept of quantum conditional matrices to  unify the solution for classical and quantum (latent) common cause problem in a principled way.\\
    \noindent$\bullet$    We evaluate the proposed approach for classical causal inference. By leveraging optimization over density matrices, the proposed approach is shown to outperform the results of classical causal inference in \citep{KocaogluNEURIPS2020} for Tubingen database \citep{Mooij2016tubingen} in section \ref{sec:evaluation}.
    
    \noindent$\bullet$    We put forth an experimental scheme that can be used to confront our theoretical framework. We consider a minimalistic model of an unknown message (possibly encrypted) with unknown origin in a two-node quantum network with the possibility of the presence of a latent common cause, where nodes are a \textit{coexisting set} of quantum systems for which a joint density matrix can be defined. Entangled quantum subsystems are used, where  subsystems are communicated over noisy channels (e.g., optical fiber)  to create such coexisting set of quantum systems.   We show that only using the joint density matrix of the observed two quantum systems, we can identify the originator of the message (i.e., the sub-system that did not encounter the noisy channel). To verify the validation of the proposed method, called \qinferg, we use realistic quantum noisy links such as quantum symmetric channel and depolarizing channel (valid for quantum networking and quantum communications) (Section \ref{sec:eval_quantum}).
    
    The rest of the paper is organized as follows. In Section \ref{sec:classic}, we review the classical causal inference approach proposed in \citep{KocaogluNEURIPS2020} for the identification of causal structures in the presence of hidden common causes. In Section \ref{sec:method}, we generalize the classical approach to the quantum domain. In Section \ref{sec:evaluation} and \ref{sec:eval_quantum}, we put forward an experimental scheme that can be used to validate our proposed approach using a minimalistic model of an unknown message (possibly encrypted) with unknown origin in a two-node/three-node classical/quantum network, respectively. In Section \ref{sec:mapping}, we explain and show why should we not map quantum to classical directly. Also, we evaluate the performance of \qinferg~on the real dataset (section \ref{sec:realdata}) with cause-effect pairs \citep{Mooij2016tubingen}, and  show that   \qinferg~outperforms in identification of latent confounders as compared to the classical approach. 
    In Section \ref{sec:mapping}, we explain and show why should we not map quantum to classical directly.

\section{Review of Classical 
Causal Inference Framework in \cite{KocaogluNEURIPS2020}}\label{sec:classic}
\begin{figure}[ht]
    \centering
    \includegraphics[width=.45\textwidth]{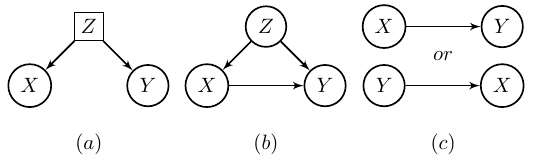}
    \caption{(a) Latent Graph, (b) Triangle Graph, and (c) Direct Graph.}
    \label{fig:possibilities}
\end{figure}

In this section, we briefly review the proposed approach in \citep{KocaogluNEURIPS2020} for confounder discovery via solving an optimization problem that its aim is to discover the trade-off between the entropy of the latent variable and the conditional mutual information of the observed variables.
Consider that the  joint distribution $P(X,Y)$ between two observed variables is given. The goal is to find a random variable $Z$  that makes $X$ and $Y$ conditionally independent given $Z$. Possible cases that can represent this situation is shown in Figure \ref{fig:possibilities}.

\begin{algorithm}[!ht]
\caption{\textbf{LatentSearch} \citep{KocaogluNEURIPS2020}}\label{alg:LatentSearch}
    \SetAlgoLined
	\footnotesize\KwIn{Supports of $X,Y$, and $Z$, respectively; Joint probability distribution $p(x,y)$; Number of iterations $N$; $\beta$ in the loss function $L=I(X; Y\text{\textbar}Z)+\beta H(Z)$, Initialization of $q_1(z\text{\textbar}x,y)$.}
	\KwOut{Joint distribution $q(x, y, z)$.}
	\For{$i = 1:N$}{
	\tcc{Form the joint distribution:}
	$q_i(x,y,z)\gets q_i(z\text{\textbar}x,y)p(x,y), \forall x,y,z$\;
	\textbf{Calculate}:
	$$q_i(z\text{\textbar}x)\gets \frac{\sum_{y\in Y} q_i(x,y,z)}{\sum_{y\in Y,z\in Z} q_i(x,y,z)},$$
	$$q_i(z\text{\textbar}y)\gets \frac{\sum_{x\in X} q_i(x,y,z)}{\sum_{x\in X,z\in Z} q_i(x,y,z)},$$ $$q_i(z)\gets \sum_{x\in X,y\in Y} q_i(x,y,z)$$
	
	\textbf{Update}:
	$$q_{i+1}(z\text{\textbar}x,y)\gets \frac{1}{F(x,y)}\frac{q_i(z\text{\textbar}x)q_i(z\text{\textbar}y)}{q_i(z)^{1-\beta}}, \textrm{ where } $$    $$F(x,y)=\sum_{z\in Z}\frac{q_i(z\text{\textbar}x)q_i(z\text{\textbar}y)}{q_i(z)^{1-\beta}}$$
	}
	\textbf{return} $q(x,y,z):= q_{N+1}(z\text{\textbar}x,y)p(x,y)$.
\end{algorithm}

In the classical causal inference, \cite{KocaogluNEURIPS2020}  distinguished between  \textit{latent graph} in Figure \ref{fig:possibilities}(a) from others in Figure \ref{fig:possibilities} based on unmeasured confounder having low Shannon entropy under certain assumptions. Formally, the following was assumed:

\begin{assumption}\label{mainassump1}
Consider any causal model with observed variables $X$ and $Y$. Let $Z$ represents the
variable that captures all latent confounders between $X$ and $Y$. Then $H(Z)<\theta,\footnote{{$\theta$ is the entropy threshold. The true $H(Z)$ is not available in practice. As discussed in \citep{KocaogluNEURIPS2020} $H(Z)$ is lower-bounded by the entropies of $X$ and $Y$, up to a scaling by a constant. For example, as suggested in \citep{KocaogluNEURIPS2020}, $\theta$ is set to $0.8\min\{H(X),H(Y)\}$ in experiments.}}$ where $H(Z) = -\sum_{i=1}^n P(x_i)\log(P(x_i))$.
\end{assumption}

Note that $I(X;Y\text{\textbar}Z)=0$ means that $Z$ makes  the variables $X$ and $Y$ conditionally independent, i.e., $X\perp\!\!\!\perp Y\text{\textbar}Z$.\footnote{Note that this is different from the notion of \textit{causal independence}, which refers to the situation where multiple causes contribute independently to a common effect \citep{zhang1996exploiting}.}  To identify latent graphs,  \cite{KocaogluNEURIPS2020} proposed an iterative algorithm (Algorithm \ref{alg:LatentSearch}) that discovers the trade-off between the entropy of the unmeasured confounder and the conditional mutual information of the observed variables. This trade-off is formally defined as follows:
\begin{equation}\label{eq:lossfunc}
    L = I(X;Y\text{\textbar}Z)+\beta H(Z)
\end{equation}

In fact, \textbf{LatentSearch} (Algorithm \ref{alg:LatentSearch}) sets $q(x,y,z) =q(z\text{\textbar}x,y)p(x,y)$ and searches over $q(z\text{\textbar}x,y)$ to find the stationary point of the loss function $L$ in Equation (\ref{eq:lossfunc}). For this purpose, \textbf{LatentSearch} returns a joint probability distribution $q(X,Y,Z)$ from which the Shannon entropy of the latent variable $W$, i.e., $H(W)$ can be computed. To verify  whether the causal graph $G=(V=\{X,Y\},E)$ is a latent graph or not, \textbf{InferGraph} (Algorithm \ref{alg:InferGraph}) \citep{KocaogluNEURIPS2020} runs \textbf{LatentSearch} multiple times and selects the
smallest $H(W)$ discovered by the algorithm among those that ensure the conditional independence of $X$ and $Y$ given $W$, i.e., $I(X;Y\text{\textbar}W)\le T$ for a practical threshold ( as suggested in \citep{KocaogluNEURIPS2020}, $T=0.001$). We refer readers to \citep{KocaogluNEURIPS2020} for more experimental settings. \cite{KocaogluNEURIPS2020} conjecture that, under Assumption \ref{mainassump1}\footnote{Note that in \citep{KocaogluNEURIPS2020} to distinguish the latent graph in Figure \ref{fig:possibilities}(a) from mediator graphs (i.e., $X\to M\to Y$, where $M$ is a latent variable), the following  is also assumed: Consider a causal model where $X$ causes $Y$. If $X$ causes $Y$ only through a
latent mediator $Z$, i.e., $X\to Z\to Y$, then $H(Z)\ge\theta$. In this work, we only focus on distinguishing between  latent graphs and direct/triangle graphs.}, and in practice, the Shannon entropy of observed variables $X$ and $Y$ for directed graphs and triangle graphs is lower-bounded by {Shannon entropy} of $X$ and $Y$, up to a scaling by a constant (as suggested in \citep{KocaogluNEURIPS2020}, $\theta=0.8\min\{H(X),H(Y)\}$). For more detailed discussion see \citep{KocaogluNEURIPS2020}.

\begin{algorithm}[!ht]
\caption{\textbf{InferGraph}: Identifying the Latent Graph \citep{KocaogluNEURIPS2020}}\label{alg:InferGraph}
    \SetAlgoLined
	\footnotesize\KwIn{Joint probability distribution $p(x,y)$; Number of iterations $N$; $I(X; Y\text{\textbar}Z)$ threshold $T$; $H(Z)$ threshold that is determined by $\theta=\alpha\min(H(X),H(Y))$; $\{\beta_i\}_{i=1}^N$; Support size of $X,Y$, and $Z$, i.e., $r,m$, and $n$, respectively.}
	\KwOut{"Latent Graph" if $Z$ is an unmeasured confounder for $X$ and $Y$, otherwise, returns "Triangle or Direct Graph".}
	\For{$i = 1:N$}{
	$q_i(x,y,z)\gets \textrm{\textbf{LatentSearch}}(p(x,y), \alpha,\beta_i,r,m,n)$\;
	Calculate $I_i(X; Y\text{\textbar}Z)$ and $H_i(Z)$ from $q_i(x,y,z)$\;
	}
	$S=\{i: I_i(X; Y\text{\textbar}Z)\le T\}$\;
	\uIf{$\min(H_i(Z):i\in S)>\theta$ or $S=\text{\O}$ }{
	\textbf{return} Triangle or Direct Graph\;
	}\Else{
	\textbf{return} Latent Graph\;
	}
\end{algorithm}

\section{Proposed Entropic Approach for Confounder Discovery in Quantum Systems}\label{sec:method}
In this section,  we provide an approach for  identifying latent graphs in quantum systems, where we assume the Assumption \ref{mainassump1}, with the entropy replaced by the von-Neumann entropy, $S(X)  = - {\sf tr}(\rho_X\log \rho_X)$.  We first briefly review the formalism of {quantum  density matrices}, which provides a solid framework for adapting classical iterative
algorithms (Algorithm \ref{alg:LatentSearch} and \ref{alg:InferGraph}) to the quantum domain. Then, the proposed algorithm to identify latent graphs is described.

\subsection{Overview of Quantum Computing }\label{sec:conditionaldensity}
Quantum theory can be understood as a non-commutative generalization of classical probability theory wherein probability measures are replaced by density
operators \citep{leifer2013towards}. The density matrix describes the quantum state of a physical system, and  allows for the calculation of the probabilities of the outcomes of any measurement performed upon this system. The density matrix is a  positive semi-definite, Hermitian matrix of trace one.  The density matrix can be written as $ \sum _{j}p_{j}\text{\textbar}\psi _{j}\rangle \langle \psi _{j}\text{\textbar}$ for some states $ \text{\textbar}\psi _{j}\rangle $  and coefficients $p_{j}$ that are non-negative and add up to one. As a generalization of classical probabilities, the density matrix corresponding to a probability distribution can be obtained where $p_j$ corresponds to the probability that the random variable is $j$ and the state $\text{\textbar}\psi _{j}\rangle$ is given as a column vector with 1 at $j^{\text{th}}$ element and zero otherwise\footnote{Note that this is not a unique method of relating the classical probabilities to quantum density matrix \citep{bradley2020language}.}.   Analogies between the classical theory of Bayesian inference and the conditional states formalism for quantum theory are listed in Table \ref{table:analogies}.

\begin{table*}[!ht]
\caption{Analogies between classical and quantum formalism}
\centering 
\begin{tabular}{|l | l|} 
\hline 
\textbf{Classical Probability} &   \textbf{Quantum Theory}\\ 
\hline \hline
probability distribution $p(X)$ &  density operator (matrix) $\rho_X$\\
\hline 
joint distribution $p(X,Y)$&  joint density $\rho_{XY}$\\
\hline
marginal distribution $p(X)=\sum_Yp(X,Y)$& partial trace $\rho_X=Tr_Y(\rho_{XY})$\\
\hline
conditional probability& conditional density matrix \\
$p(Y\text{\textbar}X)=p(X,Y)/p(X)$&$\rho_{Y\text{\textbar}X}=(\rho_X^{-1/2}\otimes I_Y)\rho_{XY}(\rho_X^{-1/2}\otimes I_Y)$\\
\hline
\end{tabular}
\label{table:analogies} 
\end{table*}

Quantum conditional densities are a generalization of classical conditional probability distributions. However, to generalize conditional probabilities to the quantum case, several  approaches have been proposed in the literature. The three following generalizations are the best known in the literature of quantum information: (1) quantum conditional expectation \citep{umegaki1962conditional}, (2) quantum conditional amplitude operator \citep{cerf1997negative,cerf1999quantum}, and (3) quantum conditional states \citep{leifer2007conditional,leifer2013towards}.  Arguably, quantum conditional states are the most useful generalization
of conditional probability from the point of view of practical applications. For example, quantum conditional states have been used in \citep{leifer2013towards} to build a quantum theory of Bayesian inference. Since quantum conditional states provides a closer analogy between quantum theory and
classical probability theory, we choose this formalism to define quantum conditional density matrices. We will see that this formalism plays a significant role in the design and success of our entropic quantum causal inference algorithm.

Following \citep{leifer2007conditional,leifer2013towards}, the conditional density matrix of $X$ given $Y$ is defined as follows:
\begin{equation*}\rho_{X\text{\textbar}Y}=(\rho_Y^{-1/2}\otimes I_X)\rho_{XY}(\rho_Y^{-1/2}\otimes I_X).\end{equation*}

Also, note that this relates the conditional density matrix and the joint density matrix, and thus the joint density matrix can also be written as 

\begin{equation*}\rho_{XY}=(\rho_Y^{1/2}\otimes I_X)\rho_{X\text{\textbar}Y}(\rho_Y^{1/2}\otimes I_X) = (\rho_X^{1/2}\otimes I_Y)\rho_{Y\text{\textbar}X}(\rho_X^{1/2}\otimes I_Y).\end{equation*}

\subsection{\qlsearch: An Algorithm for Computing Exact Quantum Common Entropy}
In this section, we propose an iterative
algorithm (Algorithm \ref{alg:qLatentSearch}) that discovers the trade-off between the entropy of the unmeasured confounder and the quantum conditional mutual information of two observed quantum systems given the unmeasured confounder. This is fundamental for designing an algorithm for the identification of latent confounders in quantum systems, as we show in the next subsection. This trade-off is formally defined as follows:
\begin{equation}\label{eq:qlossfunc}
    L = I_Q(X;Y\text{\textbar}Z)+\beta S(Z)
\end{equation}
Note that $I_Q(X;Y\text{\textbar}Z)=0$ implies that the quantum conditional independence of $X$ and $Y$ given $Z$ \cite[Theorem 3]{allen2017quantum}. Having low von Neumann entropy of hidden common cause $Z$, i.e., $S(Z)$ under the quantum version of Assumption \ref{mainassump1} enable us to identify latent graphs from direct/mediator graphs in practice, as we show in section \ref{sec:evaluation}. For this purpose, rather than searching over $\rho_{XYZ}$ and enforcing the constraint $\rho_{XY}=Tr_Z(\rho_{XYZ})$, we can search over $\rho(Z\text{\textbar}X,Y)$ and set $$\rho_{XYZ} = (\rho_{XY}^{1/2}\otimes I_Z)\rho(Z\text{\textbar}X,Y)(\rho_{XY}^{1/2}\otimes I_Z)$$ because:
\begin{equation*} \label{eq1}
\begin{split}
L &= I_Q(X;Y\text{\textbar}Z)+\beta S(Z)\\
 & = S(XZ) + S(YZ) - S(Z) - S(XYZ) + \beta S(Z)\\
 &=  S(XZ) + S(YZ) - S(XYZ) + (\beta-1) S(Z)\\
 &= S(X) + S(Z\text{\textbar}X)+S(Y)+S(Z\text{\textbar}Y)-S(XY)\\
 & -S(Z\text{\textbar}X,Y)+(\beta-1) S(Z)\\
 &= S(Z\text{\textbar}X)+S(Z\text{\textbar}Y)-S(Z\text{\textbar}X,Y)\\& +(\beta-1) S(Z)+I_Q(X;Y) 
\end{split}
\end{equation*}

Note that
$\rho(Z\text{\textbar}Y)=Tr_X((\rho^{1/2}(X\text{\textbar}Y)\otimes I_Z)\rho(Z\text{\textbar}X,Y)(\rho^{1/2}(X\text{\textbar}Y)\otimes I_Z)),$ $\rho(Z\text{\textbar}X)=Tr_Y((\rho^{1/2}(Y\text{\textbar}X)\otimes I_Z)\rho(Z\text{\textbar}X,Y)(\rho^{1/2}(Y\text{\textbar}X)\otimes I_Z)),$ and $\rho_Z=Tr_{X,Y}((\rho^{1/2}_{XY}\otimes I_Z)\rho(Z\text{\textbar}X,Y)(\rho^{1/2}_{XY}\otimes I_Z))$. So, we have $L=L(\rho(Z\text{\textbar}X,Y))$, which is the counterpart of the classical loss function in Equation \ref{eq:lossfunc} with the following differences: (i) rather than using (conditional) probability distributions, we use (conditional) density matrices, and (ii) rather than using R\'{e}nyi entropy, we use the von Neumann entropy.  

We aim to optimize the objective $L$ over $\rho(Z\text{\textbar}X,Y)$.  Although first order methods (e.g., gradient descent) or genetic algorithm (GA)\footnote{Genetic algorithm (GA) is a metaheuristic method inspired by the process of natural selection.} can be used to find a stationary point of the optimization problem in (\ref{eq:qlossfunc}),
as we empirically observed the convergence is unattainable/slow and the performance is very sensitive to the tuning parameters such as step size and the mutation probability. This optimization problem is difficult to perform numerically because the boundary of the space of positive semidefinite matrices is hard to compute. In order to provide a scalable algorithm for this optimization, we extend the iterative algorithm that was proposed for classical version of the problem in \citep{KocaogluNEURIPS2020}. 

The proposed iterative algorithm for the optimization of $L$ is described in Algorithm \ref{alg:qLatentSearch}, and is called \qlsearch. This algorithm  starts from a random initialization $\rho_1(Z\text{\textbar}X,Y)$, and then at each iteration $i$ does the following two phases to update $\rho_{i+1}(Z\text{\textbar}X,Y)$ from $\rho_i(Z\text{\textbar}X,Y)$ to finally minimize the loss function $L$ in (\ref{eq:qlossfunc}):
\begin{itemize}
    \item \textbf{Calculate Phase:} In this phase we use partial trace to get $\rho_i(Z\text{\textbar}X)$ (line 3-5), $\rho_i(Z\text{\textbar}Y)$  (line 6-8), and $\rho_Z^i$  (line 9) from $\rho_{XYZ}^i$. 
    \item \textbf{Update Phase:} In this phase we update $\rho_{i+1}(Z\text{\textbar}X,Y)$ to get $\rho_{XYZ}^{i+1}$ (line 10) for the next iteration.  
\end{itemize}

\begin{algorithm}[!ht]
\caption{\textbf{\qlsearch}, An Iterative Algorithm for Computing Exact Quantum Common Entropy}\label{alg:qLatentSearch}
    \SetAlgoLined
	\footnotesize\KwIn{Joint density matrix $\rho_{XY}$; Number of iterations $N$; $\beta$ parameter in the loss function $L=I_Q(X; Y\text{\textbar}Z)+\beta S(Z)$, Initialization of $\rho_1(Z\text{\textbar}X,Y)$.}
	\KwOut{Joint density matrix $\rho_{XYZ}$.}
	\For{$i = 1:N$}{
	\tcc{Form the joint density matrix:}
	$\rho_{XYZ}^i = (\rho_{XY}^{1/2}\otimes I_Z)\rho_i(Z\text{\textbar}X,Y)(\rho_{XY}^{1/2}\otimes I_Z)$\;
	\tcc{{Calculate Phase:}}
	\tcc{{(i) Calculate $\rho_i(Z\text{\textbar}X)$:}}
	$\rho_{XZ}^i=Tr_Y(\rho_{XYZ}^i)$
	\tcp{Then, compute $\rho_{XI_YZ}^i$ by reordering the entries of $\rho_{XZ}^i$}
	$\rho_{X}^i=Tr_Z(\rho_{XZ}^i)$\;
	$\rho_i(Z\text{\textbar}X)\gets ((\rho_{X}^i)^{-1/2}\otimes I_{YZ})\rho_{XI_YZ}^i((\rho_{X}^i)^{-1/2}\otimes I_{YZ})$\;
	\tcc{{(ii) Calculate $\rho_i(Z\text{\textbar}Y)$:}}
	$\rho_{YZ}^i=Tr_X(\rho_{XYZ}^i)$ \tcp{Then, compute $\rho_{I_XYZ}^i=I_X\otimes\rho_{YZ}^i$}
	$\rho_{Y}^i=Tr_Z(\rho_{YZ}^i)$\;
	$\rho_i(Z\text{\textbar}Y)\gets (I_{X}\otimes(\rho_{Y}^i)^{-1/2}\otimes I_{Z})\rho_{I_XYZ}^i(I_{X}\otimes(\rho_{Y}^i)^{-1/2}\otimes I_{Z})$\;
	\tcc{{(iii) Calculate $\rho_Z^i$:}}
	$\rho_{Z}^i=Tr_{XY}(\rho_{XYZ}^i)$\;
	\tcc{{Update Phase:}}
	$\rho_{i+1}(Z\text{\textbar}X,Y)\gets\exp(\log(\rho_i(Z\text{\textbar}X))+\log(\rho_i(Z\text{\textbar}Y))+(\beta-1)\log(\rho_{Z}^i))$\;
	}
	\textbf{return} $\rho_{XYZ}:=(\rho_{XY}^{1/2}\otimes I_Z)\rho_{N+1}(Z\text{\textbar}X,Y)(\rho_{XY}^{1/2}\otimes I_Z)$.
\end{algorithm}

Formally, to prove the correctness of \qlsearch, the following theorem shows that \qlsearch~converges to a stationary point of the loss function $L$ in Equation \ref{eq:qlossfunc}. The proof is available at Appendix \ref{sec:appA}. 

\begin{theorem}[Correctness of \qlsearch]\label{thm:stationary}
The stationary points of the algorithm \qlsearch~are also stationary points of the loss function $L$ in Equation \ref{eq:qlossfunc} for $0<\beta<1$.
\end{theorem}

\subsection{\textbf{\qinferg:} An Algorithm for the Identification of Latent Confounders}
In this section, we propose a quantum entropic approach to causal inference
that can discern the difference between causation and correlation. Specifically, under  Assumption \ref{mainassump1}, extended to quantum, Algorithm \ref{alg:qLatentSearch} can be used to distinguish causation from spurious correlation between two observed quantum systems. This enables us to distinguish latent graph in Figure \ref{fig:possibilities}(a) from the triangle or direct graphs in Figure \ref{fig:possibilities}(b)-(c). Our main assumption is that the latent confounders, if they exist, have small von Neumann entropy. Formally, we have:
\begin{assumption}\label{mainassump1q}
Consider any causal model with observed quantum subsystems $X$ and $Y$. Let $Z$ represents the
quantum system that captures all latent confounders between $X$ and $Y$. Then $S(Z)<\theta$, where $S(Z) = -\textrm{tr}(\rho\log\rho)$.
\end{assumption}

In other words, in Figure  \ref{fig:possibilities}(a), $S(Z)\le \theta$ for some $\theta$. Similar to the classical version of this problem, we conjecture that $\theta=\alpha\min\{S(X),S(Y)\}$ for some $\alpha < 1$. Considering Assumption \ref{mainassump1q} along with \qlsearch~(Algorithm \ref{alg:qLatentSearch}), we propose an algorithm, called \qinferg~(Algorithm \ref{alg:qInferGraph}), to identify latent graphs. 

\begin{algorithm}[!ht]
\caption{\textbf{\qinferg}: Identifying the Latent Graph}\label{alg:qInferGraph}
    \SetAlgoLined
	\footnotesize\KwIn{Joint density matrix $\rho_{XY}$; Number of iterations $N$; $I_Q(X; Y\text{\textbar}Z)$ threshold $T$; $S(Z)$ threshold that is determined by $\theta=\alpha\min(S(X),S(Y))$; $\{\beta_i\}_{i=1}^N$; The number of rows (or equivalently, columns) of $X,Y$, and $Z$, i.e., $r,m$, and $n$, respectively.}
	\KwOut{"Latent Graph" if $Z$ is an unmeasured confounder for $X$ and $Y$, otherwise, returns "Triangle or Direct Graph".}
	\For{$i = 1:N$}{
	$\rho_{XYZ}^i\gets \textrm{\textbf{\qlsearch}}(\rho_{XY}, \alpha,\beta_i,r,m,n)$\;
	Calculate $I_Q^i(X; Y\text{\textbar}Z)$ and $S_i(Z)$ from $\rho_{XYZ}^i$\;
	}
	$S=\{i: I_Q^i(X; Y\text{\textbar}Z)\le T\}$\;
	\uIf{$\min(S_i(Z):i\in S)>\theta$ or $S=\text{\O}$ }{
	\textbf{return} Triangle or Direct Graph\;
	}\Else{
	\textbf{return} Latent Graph\;
	}
\end{algorithm}

In short, \qinferg~calls \qlsearch~$N$ times to figure out if there exist a $W$, for which $I_Q(X;Y\text{\textbar}W)<T$, i.e., $W$ makes $X$ and $Y$ conditionally independent. Also, the von Neumann entropy of $W$ is enough small such that $S(W)<\alpha\min\{S(X),S(Y)\}$ for some $\alpha$ in practice. If there exist such a $W$, the algorithm declares $W$ is a latent confounder. In other words, latent graph represents correlation without causation relationship between observed quantum systems $X$ and $Y$. Otherwise, very likely such a $W$ that minimizes the loss function $L$ does not exist, and \qinferg~declares that a triangle graph or a direct graph represents the connection between $X$ and $Y$ better than a latent graph in this case. In the next section we conduct experiments to verify this procedure in practice.

\section{Evaluation on Causal Synthetic and Real Data}\label{sec:evaluation}

To verify the validity of our proposed algorithm, we put forward an experimental scheme that can be used to confront our theoretical framework. 
To show the effectiveness of the proposed approach in section \ref{sec:method}, we first use noisy links (section \ref{sec:synthetic}), where it is validated that the input before noise, as a latent confounder (hidden source), is the cause of the noisy outputs. {We will observe that the proposed approach helps achieve better tradeoff between $I_Q(X;Y\text{\textbar}Z)$ and $S(Z)$, thus helping reduce thresholds as compared to the classical approach. Using the parameter choices based on this study,  we evaluate the performance of \qinferg~on the real dataset (section \ref{sec:realdata}) with cause-effect pairs \citep{Mooij2016tubingen}, and  show that   \qinferg~ outperforms in distinguishing latent graphs from direct or triangle graphs (see Figure \ref{fig:possibilities}) as compared to the classical approach.}

\subsection{Identification of Latent Graphs in Noisy Channels}\label{sec:synthetic}
We first apply the proposed approach to a classical setup,   where two bits are transmitted over a binary symmetric channel (to illustrate the case of no confounder), or two bits are transmitted over two separate channels (to illustrate the  case of latent confounder). We show that the proposed approach outperforms the classical causal inference in \citep{KocaogluNEURIPS2020} due to the use of quantum density matrix. Finding the optima over a quantum density matrix rather than over a probability distribution provides larger degrees of freedom thus resulting in improved results. {Our results indicate that the proposed approach helps achieve better tradeoff between $I_Q(X;Y\text{\textbar}Z)$ and $S(Z)$ as compared to the classical approach.}

\begin{model}[Classical Symmetric Channel: Latent and Direct Graph]\label{model:classicchannel}
\emph{\textbf{Part I: Latent Graph.}}
\normalfont 
Assume a 2-bit input $Z\in \{00,01,10,11\}$. Let each bit of $Z$ be in the state 1 with probability $q$ and $1-q$ otherwise, and independent of each other. So, $p(Z=00)=(1-q)^2$, $p(Z=01)=p(Z=10)=q(1-q)$, and $p(Z=11)=q^2$. $Z$ is transmitted over a binary symmetric channel with independent bit error probability of $p_1$, and is denoted $X$. A cloned version of $Z$ is transmitted over a binary symmetric channel with independent bit error probability of $p_2$, and is denoted $Y$. The joint probability distribution of $X, Y$, and $Z$, where $Z$ is the cause of $X$ and $Y$, i.e., $X\gets Z\to Y$ can be computed as $p(X,Y,Z)=p(Z)p(X\text{\textbar}Z)p(y\text{\textbar}Z)$. For example, $p(01,10,00)= (1-q)q*p_1p_2*(1-p_1)p_2$. Then we marginalize out $Z$ to obtain the joint probability distribution for the latent graph $X\leftrightarrow Y$. Note that the corresponding joint density matrix $\rho_{XY}$ is a diagonal matrix that its diagonal entries come from the joint probability distribution $p(X,Y)$. The key reason of constructing $\rho_{XY}$ as the diagonal matrix from $p(X,Y)$ is to have the mixed states, so that the von-Neuman entropy of $\rho_{XY}$ is the same as the Shannon entropy of $p(X,Y)$.
\begin{figure}[ht]
    \centering
    \includegraphics[width=.45\linewidth]{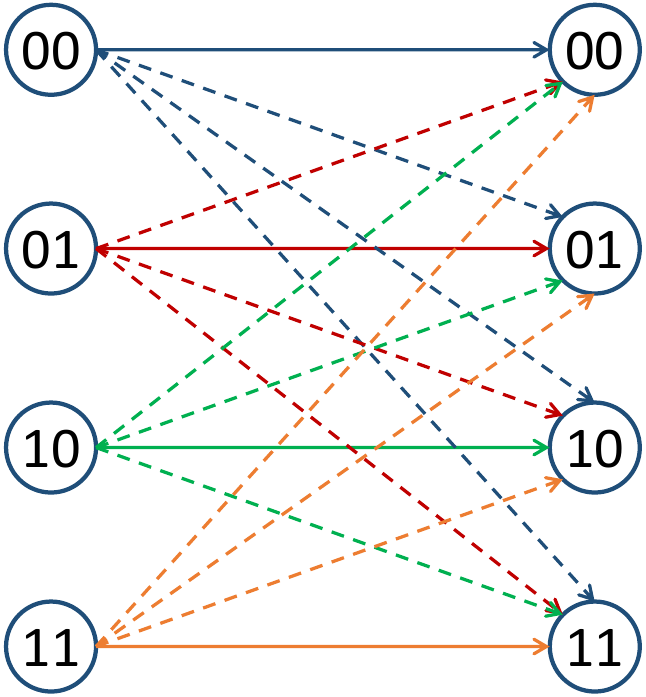}
    \caption{2-bit non-Binary symmetric channel.}
    \label{fig:NBSC}
\end{figure}

Now, we apply \qinferg~(Algorithm \ref{alg:qInferGraph}) on $\rho_{XY}$ to verify that $X$ and $Y$ are confounded by $Z$. For this purpose, we use \qlsearch~(Algorithm \ref{alg:qLatentSearch}) on {1000 different values of $\beta$, uniformly spaced in the interval $(0, 1)$.}  
We run \qlsearch~for 500 iterations each time. We use the conditional mutual information threshold of $T= 0.05$, 0.01, and 0.005. In other words, of the algorithm outputs for the 1000 $\beta$ values used, we pick the smallest entropy $W$ discovered by the algorithm among those that ensure $I(X; Y \text{\textbar}W) \le T$. Figure \ref{t:LQNBSC0.05} summarizes  the results for different $S(W)$ threshold that is determined by $\theta=\alpha\min\{S(X),S(Y)\}$ for $T=0.05$. The results for $T=0.01$ and $T=0.05$  are summarized in Figure    \ref{t:LQNBSC0.01}. For different values of $\alpha=0.2,0.3,\cdots,1$, the results are given in Figures \ref{t:LQNBSC0.05} and \ref{t:LQNBSC0.01}. We let $q=0.4$.  In each table, $\textcolor{blue}{T}$ means that \qinferg~(Algorithm \ref{alg:qInferGraph}) identifies the latent graph correctly. But, $\textcolor{red}{F}$ means that the algorithm fails to identify the latent graph. For very small or very large $p_i$'s, identification of latent graphs is difficult, while the proposed algorithm works well in most other cases.
\begin{figure}[ht]
    \centering
    \includegraphics{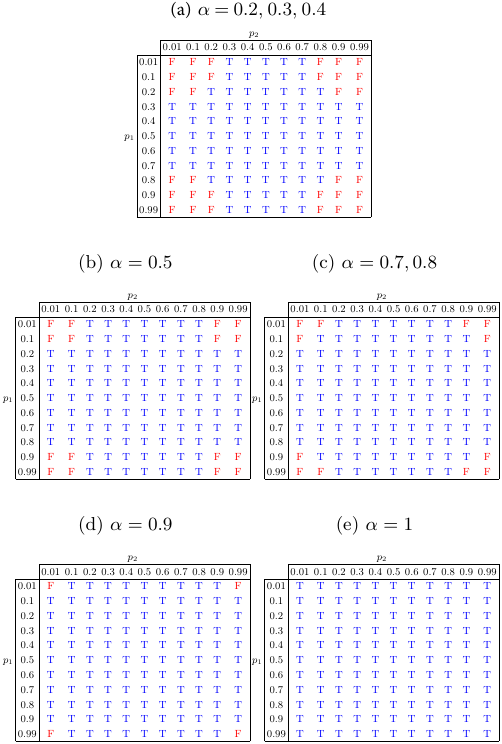}
    \caption{Validation of Latent Graph in Model \ref{model:classicchannel} (Part I) for $T=0.05$, and $\beta\in (0,1)$ via \qinferg.}\label{t:LQNBSC0.05}
\end{figure}
\begin{figure}
    \centering
    \includegraphics{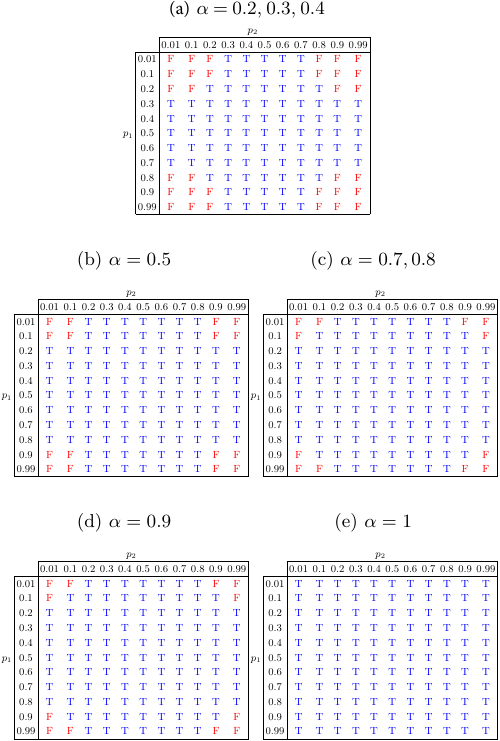}
    \caption{Validation of Latent Graph in Model \ref{model:classicchannel} (Part I) for $T= 0.01,0.005$, and $\beta\in (0,1)$ via \qinferg.}\label{t:LQNBSC0.01}
\end{figure}

Now, if we apply \textbf{InferGraph} (Algorithm \ref{alg:InferGraph}) on $p(X,Y)$ with $\alpha=0.8$, as suggested in \citep{KocaogluNEURIPS2020}, {and three more $\alpha$ parameters $\alpha=0.7, 0.9, 1$ and $\beta\in (0,1)$, we obtain the results summarized in Figure \ref{t:LNBSCclassic}.}
\begin{figure}
    \centering
    \includegraphics{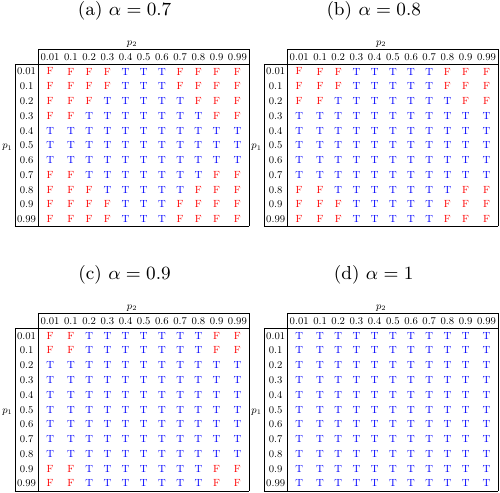}
    \caption{Validation of Latent Graph in Model \ref{model:classicchannel} (Part I) via classical causal inference (Algorithm \ref{alg:InferGraph}), $T=0.001$, and $\beta\in (0,1)$.}\label{t:LNBSCclassic}
\end{figure}

\noindent{\textbf{Some highlights for results in Part I:}}
(1) Note that when the probability of errors i.e., $p_1$ and $p_2$ are very small, the latent confounder $Z$ is hardly distinguishable from $X$ (or $Y$) and \qinferg~fails to discover the latent graph. (2) Note that \qlsearch~tries to find the stationary point(s) of the loss function $L$ in Equation (\ref{eq:qlossfunc}), and there is no guarantee to find the global optimum. {However, the performance of \qinferg~in the worst case ($\alpha=0.2,0.3,0.4$) is acceptable: true positive rate (recall) = 0.74, false positive rate (fall-out) = 0, false negative rate (miss rate) = 0.36, accuracy = 0.74.} (3) The hyperparameter $\alpha$ does not affect significantly on the quality of results in our experimental settings that indicates \qinferg~is not very sensitive to hyperparameters. {(4) It seems that the classical causal inference algorithm, i.e., \textbf{InferGraph} (Algorithm \ref{alg:InferGraph}) is much more sensitive to the choice of hyperparameter $\alpha$, while \qinferg~is more robust to the choice of this parameter. (5) The performance of  \textbf{InferGraph} (Algorithm \ref{alg:InferGraph}) for identifying latent graphs in Model \ref{model:classicchannel} (Part I) with $\alpha=0.8$ (the best $\alpha$ parameter, as suggested in \citep{KocaogluNEURIPS2020}), is the same as the performance of \qinferg~with $\alpha=0.2,0.3,0.4$. The reason is that \qinferg~constantly returns a local optima with lower entropy in comparison with the classical  \textbf{InferGraph} algorithm, because finding the optima over a quantum density matrix rather than over the probability distribution function provides larger degrees of freedom thus resulting in improved results. For example, consider the case that $p_1=0.1$ and $p_2=0.2$. Figure \ref{fig:QSCLHz2_3} shows for different points where $I_Q(X;Y\text{\textbar}Z)<T$, the values of entropy of $Z$ in a sorted order. We see that the algorithms choose lowest  entropy among these points, where \qinferg~returns 2.9 times lower local optima than \textbf{InferGraph} with entropy of 0.471543756.   Figure \ref{fig:tradeoffplaneModel1part1} shows the trade-off curve between $I_Q(X;Y\text{\textbar}Z)$ and $ S(Z)$ (respectively, between $I(X;Y\text{\textbar}Z)$ and $ H(Z)$) returned by \qlsearch~and the classical \textbf{LatentSearch} for this case that supports over observation in Figure \ref{fig:QSCLHz2_3} that the proposed approach helps achieve significantly better tradeoff. In this example, $H(X)=S(X)=1.979175042$ and $H(Y)=S(Y)=1.96290779$.} 
\begin{figure}[ht]
    \centering
    \includegraphics[width=0.75\linewidth]{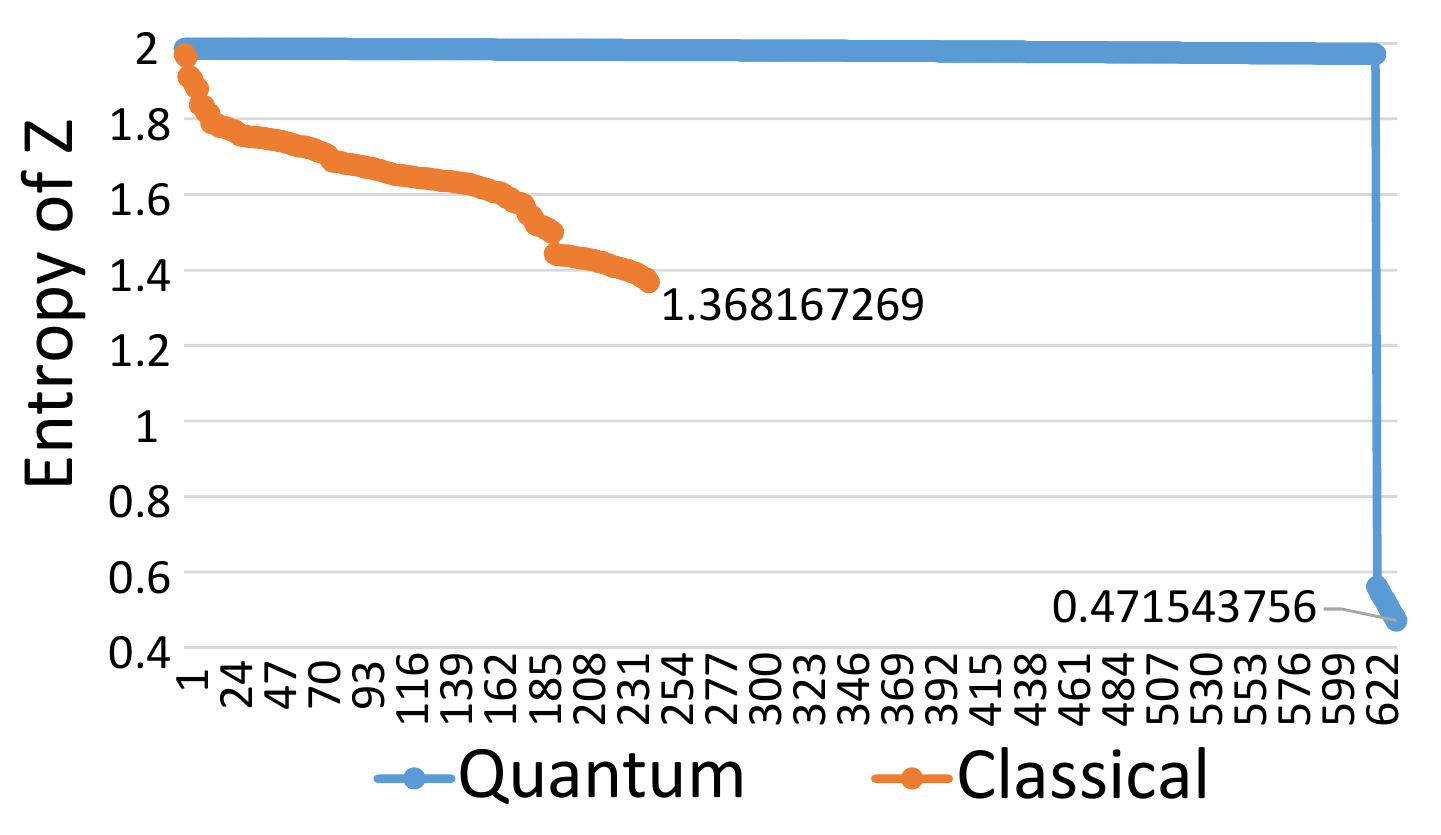}
    \caption{Entropy of possible latent confounder $Z$: \qinferg~vs the classical  \textbf{InferGraph} algorithm for Model \ref{model:classicchannel} (Part I) with $p_1=0.1$, $p_2=0.2$, and mutual conditional independence threshold  $T=0.05$.}
    \label{fig:QSCLHz2_3}
\end{figure}

\begin{figure}[ht]
\centering
\includegraphics[width=.75\linewidth]{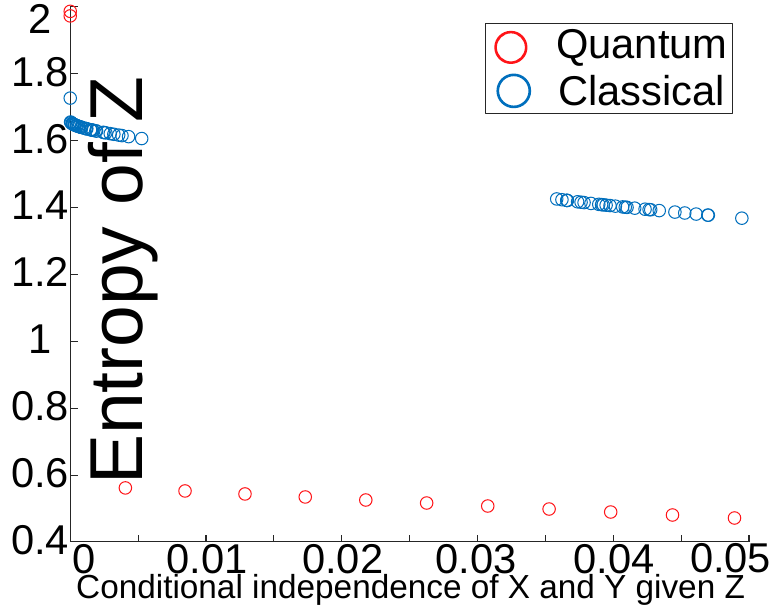}
\caption{Trade-off curve
discovered by the classical \textbf{LatentSearch} and \qlsearch~for the pair in Model \ref{model:classicchannel} (Part I) with  $p_1=0.1$ and $p_2=0.2$. Each point is the output of algorithms for a different value of $\beta\in(0,1).$}
\label{fig:tradeoffplaneModel1part1}
\end{figure}

\noindent{\textbf{Part II: Direct Graph.}} Assume that there is a 2-bit symmetric noisy channel, where there is no latent common cause, i.e., there is an input $X$ and an output $Y$, as shown in Figure \ref{fig:NBSC} with error probability $p$ on each bit, and the same properties explained in Part I. {Now, we apply \qinferg~(Algorithm \ref{alg:qInferGraph}) on $\rho_{XY}$ and \textbf{InferGraph} on $p(X,Y)$ to verify that the graph that explains the correlation between $X$ and $Y$ is a direct graph (i.e., $X\to Y$) rather than a latent graph (i.e., there exist a latent confounder $Z$ such that $X\gets Z \to Y$).}  The results of applying \qinferg~and \textbf{InferGraph}  on $\rho_{XY}$ and $p(X,Y)$ are summarized in Figures \ref{t:DQNBSC} and \ref{t:DNBSCclassic}, respectively. $\textcolor{blue}{T}$ means that \qinferg~(Algorithm \ref{alg:qInferGraph}) identifies the direct graph correctly. But, $\textcolor{red}{F}$ means that the algorithm fails to identify the direct graph.
\begin{figure}
    \centering
    \includegraphics{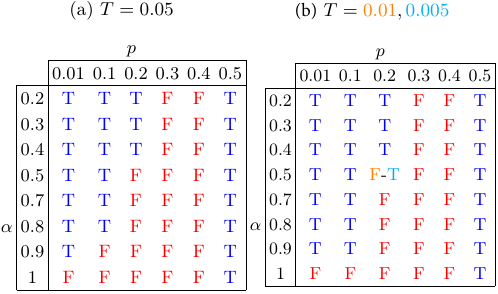}
    \caption{Validation of Direct Graph in Model   \ref{model:classicchannel} (Part II) via \qinferg, and $\beta\in (0,1)$.}
    \label{t:DQNBSC}
\end{figure}
\begin{figure}
    \centering
    \includegraphics{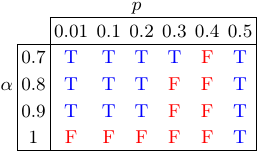}
    \caption{Validation of Latent Graph in Model \ref{model:classicchannel} (Part II) via classical causal inference (Algorithm \ref{alg:InferGraph}), and $\beta\in(0,1)$.}\label{t:DNBSCclassic}
\end{figure}
    

\noindent{\textbf{Some highlights for results in Part II:}} (1) {The performance of  \textbf{InferGraph} (Algorithm \ref{alg:InferGraph}) for identifying latent graphs in Model \ref{model:classicchannel} (Part I) with $\alpha=0.8$ (the best $\alpha$ parameter, as suggested in \citep{KocaogluNEURIPS2020}), is the same as the performance of \qinferg~with $\alpha=0.2,0.3,0.4$. This confirms our observation in Part I of this model. For example, consider the case that $p=0.2$. Figure \ref{fig:QSCDHzQvsC3} shows that for this case \qinferg~returns a better local optima than \textbf{InferGraph} with entropy of 0.864236474. Figure \ref{fig:tradeoffplaneModel1part2} shows the trade-off curve returned by the classical \textbf{LatentSearch} and \qlsearch~for this case that supports over observation in Figure \ref{fig:QSCLHz2_3}. In this example, $H(X)=S(X)=1.979175042$ and $H(Y)=S(Y)=1.941901189$. (2) Although the performance of the classical algorithm (Algorithm \ref{alg:InferGraph}), where $\alpha= 0.5,0.7$, is  better than \qinferg~for Model \ref{model:classicchannel} (Part II), its performance for Model \ref{model:classicchannel} (Part I), where there is a latent confounder, is not satisfactory.}
\begin{figure}[ht]
    \centering
    \includegraphics[width=0.75\linewidth]{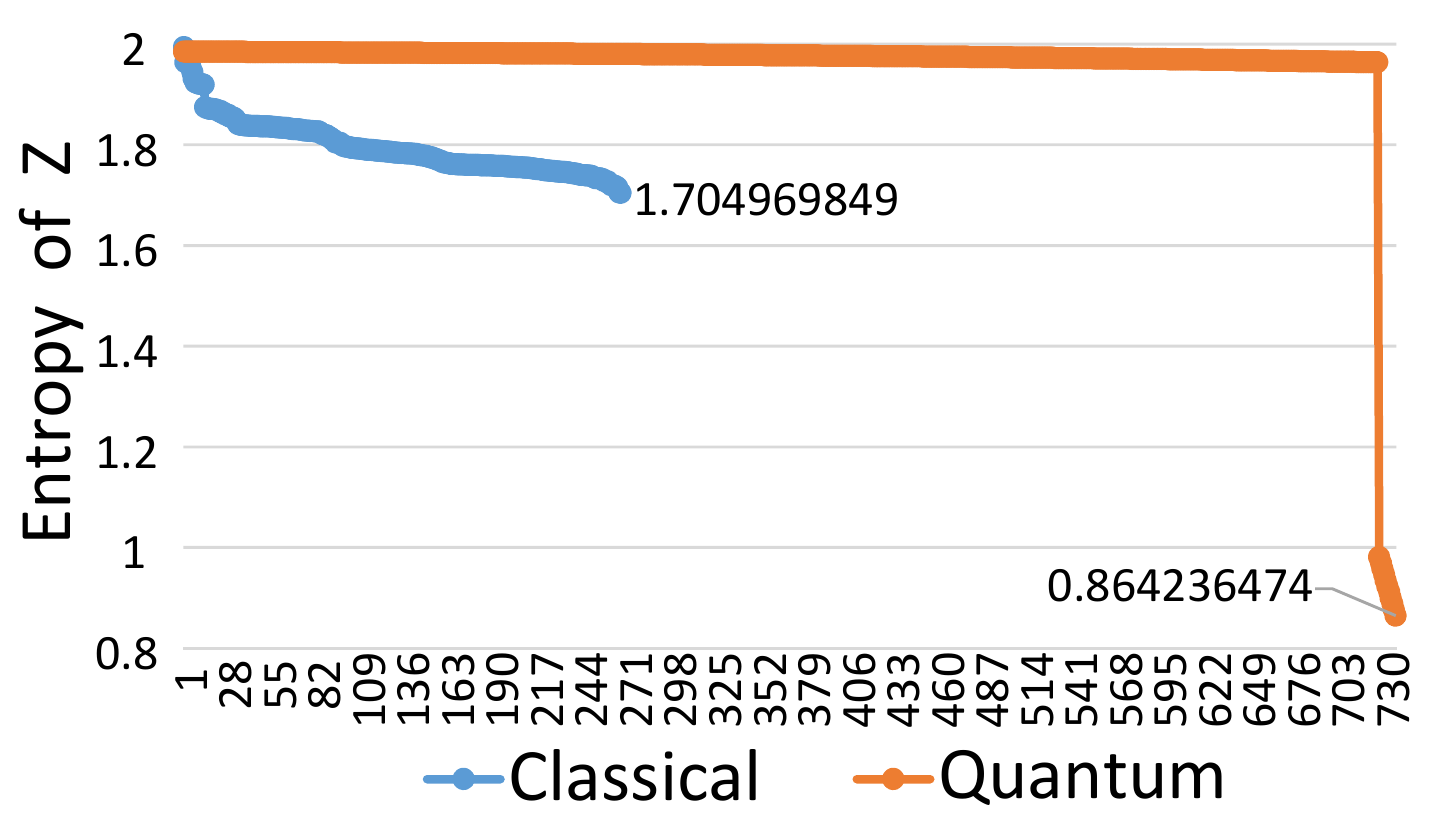}
    \caption{Entropy of possible latent confounder $Z$: \qinferg~vs the classical  \textbf{InferGraph} algorithm for Model \ref{model:classicchannel} (Part II) with $p=0.2$, and mutual conditional independence threshold  $T=0.05$.}
    \label{fig:QSCDHzQvsC3}
\end{figure}
\begin{figure}[ht]
    \centering
\includegraphics[width=.75\linewidth]{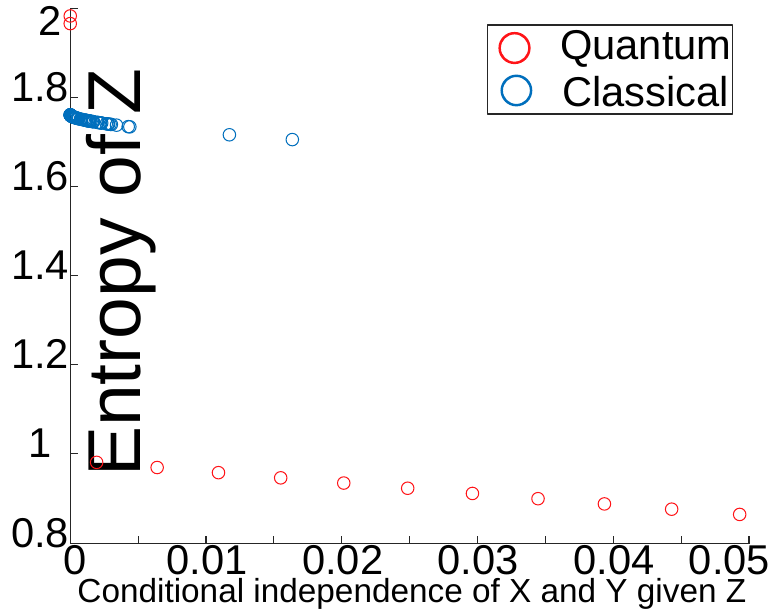}
\caption{Trade-off curve
discovered by the classical \textbf{LatentSearch} and \qlsearch~for the pair in Model \ref{model:classicchannel} (Part II) with  $p=0.2$. Each point is the output of algorithms for a different value of $\beta\in(0,1).$}
\label{fig:tradeoffplaneModel1part2}
\end{figure}

In conclusion, results from Part I and II, indicate that \qinferg~is a more consistent and less sensitive to the change of parameters than its counterpart in the classical causal inference, even for the classical data. {The proposed approach helps achieve better tradeoff curves between the two metrics.  In addition, for the classical \textbf{InferGraph} algorithm, as suggested in \citep{KocaogluNEURIPS2020}, the best hyperparameters are $\alpha=0.8$ and $T=0.001$; while for \qinferg~the best hyperparameters in this setting are $\alpha=0.2$ and $T=0.005$. As we mentioned earlier in this model, since \qinferg~consistently returns a local optima with lower entropy than \textbf{InferGraph}, we need to use a smaller $\alpha$ parameter ($\alpha=0.2$) in \qinferg.  Thus, in the remainder of the paper, we will use these parameter values.}
\end{model}

{\subsection{Distinguishing Cause from Effect Using Observational Data: Tuebingen dataset}\label{sec:realdata}
Inferring causal relationships from observational data alone is a challenging task even in the most elementary form of such a causal discovery problem, i.e., determining whether $X$ causes $Y$ or, alternatively, $Y$ causes $X$, given only joint measurements of both variables. Tuebingen dataset is a benchmark database that includes more than 100 different cause-effect pairs selected from various domains (e.g., meteorology, biology, medicine, engineering, economy, etc.) \citep{Mooij2016tubingen}. Here, we only consider the first 41 pairs of cause-effect datasets, available at : \url{https://webdav.tuebingen.mpg.de/cause-effect/}, to evaluate the performance of \qinferg~on real data. According to the website of Tubingen database, each datafile contains two variables, where one of them is the cause and the other one is the effect with the possibility of the existence of a latent confounder. So, all cause-effect pairs have a form of a direct/triangle graph, as shown in Figure \ref{fig:tubingen}.
\begin{figure}[ht]
    \centering
    \includegraphics{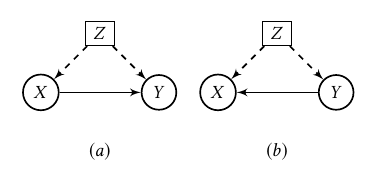}
    \caption{Tubingen: Database with cause-effect pairs of the form (a) or (b).}
    \label{fig:tubingen}
\end{figure}

For example, the first cause-effect pair from Tubingen database consists of of two variables: altitude and  temperature, where the ground truth says altitude causes temperature. Figure \ref{fig:Tubingenpair1} shows a scatter plot for this case. Note that data was taken at 349 different stations.
\begin{figure}
    \centering
    \includegraphics{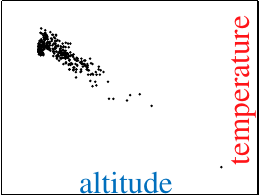}
    \caption{First cause-effect pair of data from Tubingen database: altitude causes temperature.}
    \label{fig:Tubingenpair1}
\end{figure}

Here, the goal is to decide whether the correlation between altitude and temperature is only due to a common cause (latent graph) or one of them causes the other one (direct/triangle graph). Assume that $X$ is altitude and $Y$ is temperature. \qinferg~with mutual conditional independence threshold  $T=0.05$ and $\beta\in(0,1)$ returns a $Z$ with entropy 0.644801839
which is greater than $0.2\min\{H(X),H(Y)\}=0.1582$, where $H(X)=0.791249247$ and $H(Y)=0.989477143$. This confirms that this is not a case of \textit{correlation without causation}, and very likely $X$ and $Y$ are causally related. Since there is a big gap between the threshold of $0.1582$ and the returned entropy of 0.644801839, the decision is easier. Note that to deal
with continuous variables, we discretized continuous variables with 5 levels for both $X$ and $Y$.  Also, Figure \ref{fig:P1HzQvcC} confirms our observation in section \ref{sec:synthetic} regarding finding the optima over a quantum density matrix rather than over a probability distribution where we note that the classical approach does not give any feasible point and thus does not generate any points in the figure. In fact, searching for (local) optima over a quantum density matrix provides larger degrees of freedom thus resulting in improved results. Now, we confirm this observation for the first 41 pairs of Tubingen data.

\begin{figure}[ht]
    \centering
    \includegraphics[width=0.75\linewidth]{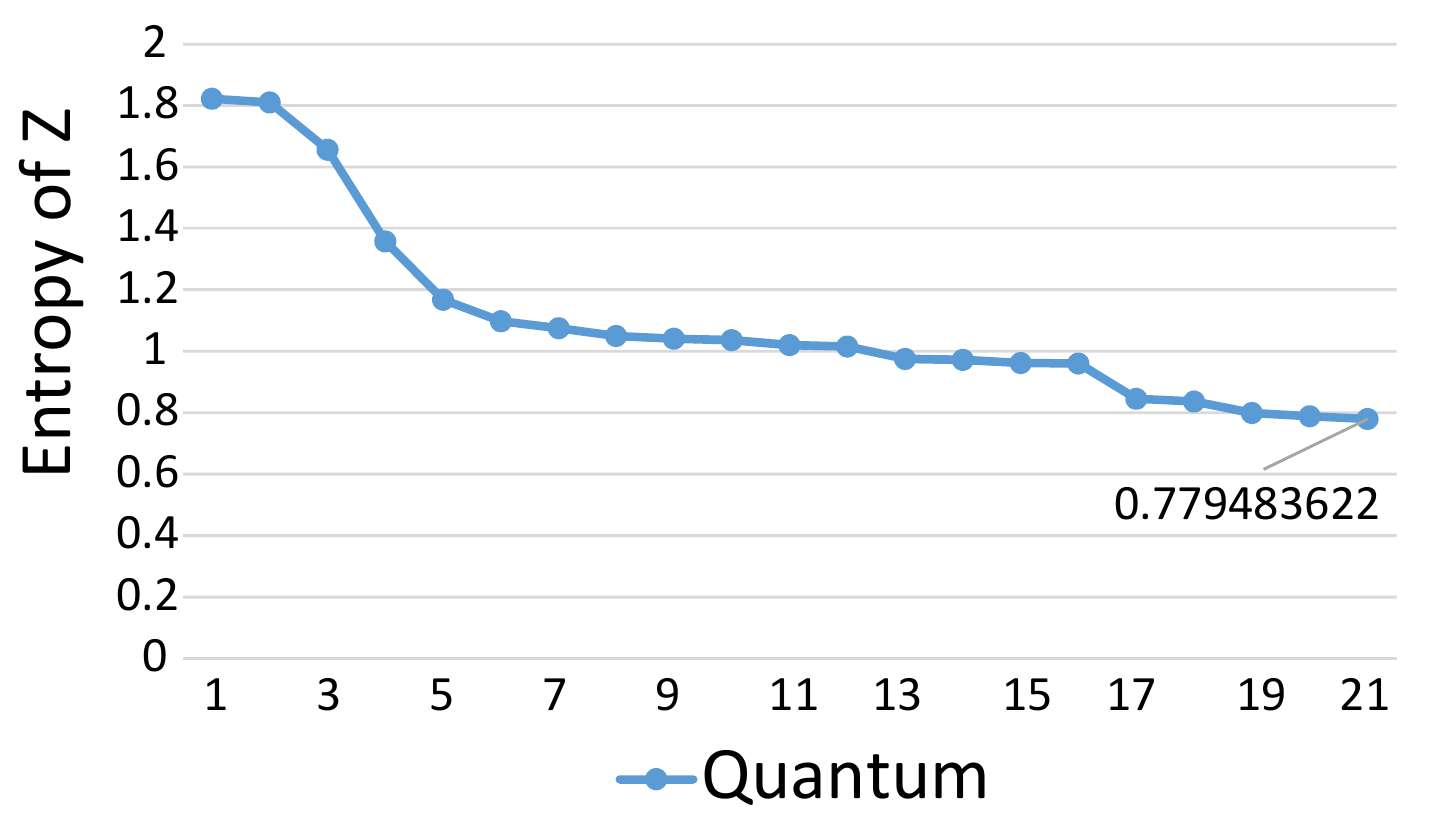}
    \caption{Entropy of possible latent confounder $Z$: \qinferg~vs the classical  \textbf{InferGraph} algorithm for the first cause-effect pair of Tubingen database with mutual conditional independence threshold  $T=0.005$ for \qinferg. Note that for the classical \textbf{InferGraph} algorithm the minimum obtained mutual conditional independence $I(X;Y\text{\textbar}Z)$ is 0.174265303 and there exist no $Z$ such that  $I(X;Y\text{\textbar}Z)< 0.001$. Also, note that $H(X)=0.791249247$ and $H(Y)=0.989477143$, where $X$ is altitude and $Y$ is temperature.}
    \label{fig:P1HzQvcC}
\end{figure}

\paragraph{Results over the first 41 pairs of Tubingen database.}
As we discussed in section \ref{sec:synthetic}, for the classical \textbf{InferGraph} algorithm, as suggested in \citep{KocaogluNEURIPS2020}, the best hyperparameters are $\alpha=0.8$ and $T=0.001$; while for \qinferg~the best hyperparameters are $\alpha=0.2$ and $T=0.005$. Table \ref{tab:Tubingen} summarizes the results for  \qinferg~and the classical \textbf{InferGraph} algorithm with the above mentioned parameters on Tubingen database.
Thus, we see that the proposed approach helps achieve significantly better accuracy (0.83) as compared to lower than 50\% in the baseline approach. In addition, the false negative rate of 0.17 in the proposed approach is significantly lower than the baselines which have this rate above 0.5. Thus, we see that the proposed approach outperforms classical approach on  Tubingen database.
\begin{table*}[ht]
    \caption{Performance of \qinferg~vs classical \textbf{InferGraph} on Tubingen database.}
    \label{tab:Tubingen}
    \begin{center}
\resizebox{\textwidth}{!}{\begin{tabular}{ |c|c|c|c| c|} 
 \hline
 Algorithm&True Positive & False Positive & False Negative& Accuracy \\
 \hline
\qinferg~($\alpha=0.2,T=0.005$) &\textbf{0.83} & 0 & \textbf{0.17}& \textbf{0.83}\\
\hline
Classical \textbf{InferGraph} ($\alpha=0.8,T=0.001$)& 0.32 & 0 & 0.68& 0.32\\ 
 \hline
Classical \textbf{InferGraph} ($\alpha=0.7,T=0.001$)& 0.49 & 0 & 0.51&0.49 \\ 
 \hline
\end{tabular}}
\end{center}
\end{table*}}

\section{Evaluation on Quantum Causal Synthetic Data}\label{sec:eval_quantum}

Since there is no quantum cause-effect repository to verify the validity of our proposed algorithm, we put forward an experimental scheme that can be used to confront our theoretical framework. 
To show the effectiveness of the proposed approach in section \ref{sec:method}, we use quantum noisy links, where it is validated that the input before noise, as a latent confounder (hidden source), is the cause of the noisy outputs.

We first apply our proposed approach on a quantum (non-classical) model, where mixed entangled quantum subsystems are used for which  subsystems are communicated over noisy channels (e.g., optical fiber)  to create a coexisting set of quantum systems.
\begin{model}[Depolarizing Quantum Channel: Latent Graph and Direct Graph]\label{model:depolar}~\\
\emph{\textbf{Part I: Latent Graph.}}
\normalfont
Assume that there are real numbers $\gamma_1$, $\gamma_2$, $\lambda_1$, and $\lambda_2$ such that $\gamma_1^2+\lambda_1^2=1$ and $\gamma_2^2+\lambda_2^2=1$. We consider a joint entangled system (of three qubits) as the mixture of the following pure density matrices:

\vspace{\baselineskip}
\noindent\resizebox{\linewidth}{!}{%
$\displaystyle
    \left\{
    \begin{array}{lll}
    [(\gamma_1 \text{\textbar}0\rangle + \lambda_1\text{\textbar}1\rangle)(\gamma_1 \text{\textbar}0\rangle + \lambda_1\text{\textbar}1\rangle)(\gamma_1 \text{\textbar}0\rangle + \lambda_1\text{\textbar}1\rangle)] [(\gamma_1 \text{\textbar}0\rangle + \lambda_1\text{\textbar}1\rangle)(\gamma_1 \text{\textbar}0\rangle + \lambda_1\text{\textbar}1\rangle)(\gamma_1 \text{\textbar}0\rangle + \lambda_1\text{\textbar}1\rangle)]^\dagger & & q\\
    
     [(\gamma_2 \text{\textbar}0\rangle + \lambda_2\text{\textbar}1\rangle)(\gamma_2 \text{\textbar}0\rangle + \lambda_2\text{\textbar}1\rangle)(\gamma_2 \text{\textbar}0\rangle + \lambda_2\text{\textbar}1\rangle)] [(\gamma_2 \text{\textbar}0\rangle + \lambda_2\text{\textbar}1\rangle)(\gamma_2 \text{\textbar}0\rangle + \lambda_2\text{\textbar}1\rangle)(\gamma_2 \text{\textbar}0\rangle + \lambda_2\text{\textbar}1\rangle)]^\dagger & & 1-q
\end{array} \right.
$}
\vspace{\baselineskip}

In other words, the system considered has density matrix $q[(\gamma_1 \text{\textbar}0\rangle + \lambda_1\text{\textbar}1\rangle)(\gamma_1 \text{\textbar}0\rangle + \lambda_1\text{\textbar}1\rangle)(\gamma_1 \text{\textbar}0\rangle + \lambda_1\text{\textbar}1\rangle)] [(\gamma_1 \text{\textbar}0\rangle + \lambda_1\text{\textbar}1\rangle)(\gamma_1 \text{\textbar}0\rangle + \lambda_1\text{\textbar}1\rangle)(\gamma_1 \text{\textbar}0\rangle + \lambda_1\text{\textbar}1\rangle)]^\dagger  + (1-q) [(\gamma_2 \text{\textbar}0\rangle + \lambda_2\text{\textbar}1\rangle)(\gamma_2 \text{\textbar}0\rangle + \lambda_2\text{\textbar}1\rangle)(\gamma_2 \text{\textbar}0\rangle + \lambda_2\text{\textbar}1\rangle)] [(\gamma_2 \text{\textbar}0\rangle + \lambda_2\text{\textbar}1\rangle)(\gamma_2 \text{\textbar}0\rangle + \lambda_2\text{\textbar}1\rangle)(\gamma_2 \text{\textbar}0\rangle + \lambda_2\text{\textbar}1\rangle)]^\dagger $.  The system is a mixture of two pure density matrices. This quantum system has entanglement among the three quantum bits. Let the second quantum bit is transmitted over a \emph{quantum depolarizing channel}  with error probability $p_1$, and the third quantum bit is transmitted over a \emph{quantum depolarizing channel} with error probability $p_2$. Note that the depolarizing channel with error probability $p$ has no error with probability $1-p$, and each of the phase-flip, bit-flip, or the combination of phase-flip and bit-flip errors with probability $p/3$ \citep{nielsen2002quantum}. With this setup, the joint density matrix is given as $\rho_{ZXY} = q \rho_{ZXY}^{\gamma_1,\lambda_1} + (1-q) \rho_{ZXY}^{\gamma_2,\lambda_2}$, where $\rho_{ZXY}^{\gamma,\lambda}$ is given as the mixture of the following pure density matrices:


\vspace{\baselineskip}
\noindent\resizebox{\linewidth}{!}{%
$\displaystyle
    \left\{
    \begin{array}{lll}
    [(\gamma \text{\textbar}0\rangle + \lambda\text{\textbar}1\rangle)(\gamma \text{\textbar}0\rangle + \lambda\text{\textbar}1\rangle)(\gamma \text{\textbar}0\rangle + \lambda\text{\textbar}1\rangle)] [(\gamma \text{\textbar}0\rangle + \lambda\text{\textbar}1\rangle)(\gamma \text{\textbar}0\rangle + \lambda\text{\textbar}1\rangle)(\gamma \text{\textbar}0\rangle + \lambda\text{\textbar}1\rangle)]^\dagger & & (1-p_1)(1-p_2)\\
    
    [(\gamma \text{\textbar}0\rangle + \lambda\text{\textbar}1\rangle)(\gamma \text{\textbar}0\rangle + \lambda\text{\textbar}1\rangle)(\gamma \text{\textbar}0\rangle - \lambda\text{\textbar}1\rangle)] [(\gamma \text{\textbar}0\rangle + \lambda\text{\textbar}1\rangle)(\gamma \text{\textbar}0\rangle + \lambda\text{\textbar}1\rangle)(\gamma \text{\textbar}0\rangle - \lambda\text{\textbar}1\rangle)]^\dagger & & (1-p_1)(p_2/3) \\
    
    [(\gamma \text{\textbar}0\rangle + \lambda\text{\textbar}1\rangle)(\gamma \text{\textbar}0\rangle + \lambda\text{\textbar}1\rangle)(\lambda \text{\textbar}0\rangle +\gamma \text{\textbar}1\rangle)] [(\gamma \text{\textbar}0\rangle + \lambda\text{\textbar}1\rangle)(\gamma \text{\textbar}0\rangle + \lambda\text{\textbar}1\rangle)(\lambda \text{\textbar}0\rangle +\gamma \text{\textbar}1\rangle)]^\dagger & &  (1-p_1)(p_2/3) \\
    
    [(\gamma \text{\textbar}0\rangle + \lambda\text{\textbar}1\rangle)(\gamma \text{\textbar}0\rangle + \lambda\text{\textbar}1\rangle)(-\lambda \text{\textbar}0\rangle +\gamma \text{\textbar}1\rangle)] [(\gamma \text{\textbar}0\rangle + \lambda\text{\textbar}1\rangle)(\gamma \text{\textbar}0\rangle + \lambda\text{\textbar}1\rangle)(-\lambda \text{\textbar}0\rangle +\gamma \text{\textbar}1\rangle)]^\dagger & &  (1-p_1)(p_2/3) \\
    
    [(\gamma \text{\textbar}0\rangle + \lambda\text{\textbar}1\rangle)(-\lambda \text{\textbar}0\rangle +\gamma \text{\textbar}1\rangle)(\gamma \text{\textbar}0\rangle + \lambda\text{\textbar}1\rangle)] [(\gamma \text{\textbar}0\rangle + \lambda\text{\textbar}1\rangle)(-\lambda \text{\textbar}0\rangle +\gamma \text{\textbar}1\rangle)(\gamma \text{\textbar}0\rangle + \lambda\text{\textbar}1\rangle)]^\dagger & & (p_1/3)(1-p_2)\\
    
    [(\gamma \text{\textbar}0\rangle + \lambda\text{\textbar}1\rangle)(-\lambda \text{\textbar}0\rangle +\gamma \text{\textbar}1\rangle)(\lambda \text{\textbar}0\rangle +\gamma \text{\textbar}1\rangle)] [(\gamma \text{\textbar}0\rangle + \lambda\text{\textbar}1\rangle)(-\lambda \text{\textbar}0\rangle +\gamma \text{\textbar}1\rangle)(\lambda \text{\textbar}0\rangle +\gamma \text{\textbar}1\rangle)]^\dagger & & (p_1/3)(p_2/3) \\
    
    [(\gamma \text{\textbar}0\rangle + \lambda\text{\textbar}1\rangle)(-\lambda \text{\textbar}0\rangle +\gamma \text{\textbar}1\rangle)(-\lambda \text{\textbar}0\rangle +\gamma \text{\textbar}1\rangle)] [(\gamma \text{\textbar}0\rangle + \lambda\text{\textbar}1\rangle)(-\lambda \text{\textbar}0\rangle +\gamma \text{\textbar}1\rangle)(-\lambda \text{\textbar}0\rangle +\gamma \text{\textbar}1\rangle)]^\dagger & & (p_1/3)(p_2/3) \\
    
    [(\gamma \text{\textbar}0\rangle + \lambda\text{\textbar}1\rangle)(-\lambda \text{\textbar}0\rangle +\gamma \text{\textbar}1\rangle)(\gamma \text{\textbar}0\rangle - \lambda\text{\textbar}1\rangle)] [(\gamma \text{\textbar}0\rangle + \lambda\text{\textbar}1\rangle)(-\lambda \text{\textbar}0\rangle +\gamma \text{\textbar}1\rangle)(\gamma \text{\textbar}0\rangle - \lambda\text{\textbar}1\rangle)]^\dagger & & (p_1/3)(p_2/3) \\
    
    [(\gamma \text{\textbar}0\rangle + \lambda\text{\textbar}1\rangle)(\lambda \text{\textbar}0\rangle +\gamma \text{\textbar}1\rangle)(\gamma \text{\textbar}0\rangle + \lambda\text{\textbar}1\rangle)] [(\gamma \text{\textbar}0\rangle + \lambda\text{\textbar}1\rangle)(\lambda \text{\textbar}0\rangle +\gamma \text{\textbar}1\rangle)(\gamma \text{\textbar}0\rangle + \lambda\text{\textbar}1\rangle)]^\dagger & & (p_1/3)(1-p_2)\\
    
    [(\gamma \text{\textbar}0\rangle + \lambda\text{\textbar}1\rangle)(\lambda \text{\textbar}0\rangle +\gamma \text{\textbar}1\rangle)(\lambda \text{\textbar}0\rangle +\gamma \text{\textbar}1\rangle)] [(\gamma \text{\textbar}0\rangle + \lambda\text{\textbar}1\rangle)(\lambda \text{\textbar}0\rangle +\gamma \text{\textbar}1\rangle)(\lambda \text{\textbar}0\rangle +\gamma \text{\textbar}1\rangle)]^\dagger & & (p_1/3)(p_2/3) \\
    
    [(\gamma \text{\textbar}0\rangle + \lambda\text{\textbar}1\rangle)(\lambda \text{\textbar}0\rangle +\gamma \text{\textbar}1\rangle)(-\lambda \text{\textbar}0\rangle +\gamma \text{\textbar}1\rangle)] [(\gamma \text{\textbar}0\rangle + \lambda\text{\textbar}1\rangle)(\lambda \text{\textbar}0\rangle +\gamma \text{\textbar}1\rangle)(-\lambda \text{\textbar}0\rangle +\gamma \text{\textbar}1\rangle)]^\dagger & & (p_1/3)(p_2/3) \\
    
    [(\gamma \text{\textbar}0\rangle + \lambda\text{\textbar}1\rangle)(\lambda \text{\textbar}0\rangle +\gamma \text{\textbar}1\rangle)(\gamma \text{\textbar}0\rangle - \lambda\text{\textbar}1\rangle)] [(\gamma \text{\textbar}0\rangle + \lambda\text{\textbar}1\rangle)(\lambda \text{\textbar}0\rangle +\gamma \text{\textbar}1\rangle)(\gamma \text{\textbar}0\rangle - \lambda\text{\textbar}1\rangle)]^\dagger & & (p_1/3)(p_2/3) \\
        
    [(\gamma \text{\textbar}0\rangle + \lambda\text{\textbar}1\rangle)(\gamma \text{\textbar}0\rangle - \lambda\text{\textbar}1\rangle)(\gamma \text{\textbar}0\rangle + \lambda\text{\textbar}1\rangle)] [(\gamma \text{\textbar}0\rangle + \lambda\text{\textbar}1\rangle)(\gamma \text{\textbar}0\rangle - \lambda\text{\textbar}1\rangle)(\gamma \text{\textbar}0\rangle + \lambda\text{\textbar}1\rangle)]^\dagger & & (p_1/3)(1-p_2)\\
    
    [(\gamma \text{\textbar}0\rangle + \lambda\text{\textbar}1\rangle)(\gamma \text{\textbar}0\rangle - \lambda\text{\textbar}1\rangle)(\lambda \text{\textbar}0\rangle +\gamma \text{\textbar}1\rangle)] [(\gamma \text{\textbar}0\rangle + \lambda\text{\textbar}1\rangle)(\gamma \text{\textbar}0\rangle - \lambda\text{\textbar}1\rangle)(\lambda \text{\textbar}0\rangle +\gamma \text{\textbar}1\rangle)]^\dagger & & (p_1/3)(p_2/3) \\
    
    [(\gamma \text{\textbar}0\rangle + \lambda\text{\textbar}1\rangle)(\gamma \text{\textbar}0\rangle - \lambda\text{\textbar}1\rangle)(-\lambda \text{\textbar}0\rangle +\gamma \text{\textbar}1\rangle)] [(\gamma \text{\textbar}0\rangle + \lambda\text{\textbar}1\rangle)(\gamma \text{\textbar}0\rangle - \lambda\text{\textbar}1\rangle)(-\lambda \text{\textbar}0\rangle +\gamma \text{\textbar}1\rangle)]^\dagger & & (p_1/3)(p_2/3) \\
    
    [(\gamma \text{\textbar}0\rangle + \lambda\text{\textbar}1\rangle)(\gamma \text{\textbar}0\rangle - \lambda\text{\textbar}1\rangle)(\gamma \text{\textbar}0\rangle - \lambda\text{\textbar}1\rangle)] [(\gamma \text{\textbar}0\rangle + \lambda\text{\textbar}1\rangle)(\gamma \text{\textbar}0\rangle - \lambda\text{\textbar}1\rangle)(\gamma \text{\textbar}0\rangle - \lambda\text{\textbar}1\rangle)]^\dagger & & (p_1/3)(p_2/3)
\end{array} \right.
$}
\vspace{\baselineskip}

We note that $X$ and $Y$ coexist, thus we can find joint density matrix of $X$ and $Y$ by tracing out $Z$ in $\rho_{ZXY}$. 
Then, we apply \qinferg~(Algorithm \ref{alg:qInferGraph}) on $\rho_{XY}$ to verify that $X$ and $Y$ are confounded by a latent confounder. For this purpose, we use the same parameters specification as explained in Model \ref{model:classicchannel} with {$\alpha=0.2, \beta\in(0,1), T=0.005$, and $q=0.4$.} Figure \ref{t:Ldepolar} summarizes the results , where $\alpha=0.2$. $\textcolor{blue}{T}$ means that \qinferg~(Algorithm \ref{alg:qInferGraph}) identifies the latent graph correctly. But, $\textcolor{red}{F}$ means that the algorithm fails to identify the latent graph.  The results confirm our observations that we made in Model \ref{model:classicchannel} (Part I). However, in this case \qinferg~has a higher performance quality. For example, for $\alpha=0.2$ we have:  true positive rate (recall) = 1, false positive rate (fall-out) = 0, false negative rate (miss rate) = 0, accuracy = 1.
\begin{figure}
    \centering
    \includegraphics{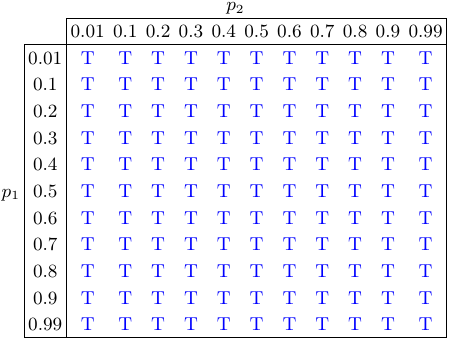}
    \caption{Validation of Latent Graph in Model \ref{model:depolar} (Part I)  for 	$\alpha=0.2$, and $\beta\in (0,1)$ via \qinferg, and  with the density matrix obtained from $0.6\rho_{ZXY}^{1/\sqrt{2},1/\sqrt{2}}+0.4\rho_{ZXY}^{0.6,0.8}$ via tracing out $Z$.}\label{t:Ldepolar}
\end{figure}

{\textbf{Part II: Direct Graph.}} Assume that there are real numbers $\gamma_1$, $\gamma_2$, $\lambda_1$, and $\lambda_2$ such that $\gamma_1^2+\lambda_1^2=1$ and $\gamma_2^2+\lambda_2^2=1$. We consider a joint entangled system (of two qubits) as the mixture of the following pure density matrices:

\vspace{\baselineskip}
\noindent\resizebox{\linewidth}{!}{%
$\displaystyle
    \left\{
    \begin{array}{lll}
   (\gamma_1^2\text{\textbar}00\rangle+\gamma_1\lambda_1\text{\textbar}01\rangle+\gamma_1\lambda_1\text{\textbar}10\rangle+\lambda_1^2\text{\textbar}11\rangle)(\gamma_1^2\text{\textbar}00\rangle+\gamma_1\lambda_1\text{\textbar}01\rangle+\gamma_1\lambda_1\text{\textbar}10\rangle+\lambda_1^2\text{\textbar}11\rangle)^\dagger & & q\\
    
     (\gamma_2^2\text{\textbar}00\rangle+\gamma_2\lambda_2\text{\textbar}01\rangle+\gamma_2\lambda_2\text{\textbar}10\rangle+\lambda_2^2\text{\textbar}11\rangle)(\gamma_2^2\text{\textbar}00\rangle+\gamma_2\lambda_2\text{\textbar}01\rangle+\gamma_2\lambda_2\text{\textbar}10\rangle+\lambda_2^2\text{\textbar}11\rangle)^\dagger & & 1-q
\end{array} \right.
$}
\vspace{\baselineskip}

The system is a mixture of two pure density matrices. This quantum system has entanglement among the two quantum bits. Let the second quantum bit is transmitted over a \emph{quantum depolarizing channel}  with error probability $p$. With this setup, the joint density matrix is given as $\rho_{XY} = q \rho_{XY}^{\gamma_1,\lambda_1} + (1-q) \rho_{XY}^{\gamma_2,\lambda_2}$, where $\rho_{XY}^{\gamma,\lambda}$ is given as the mixture of the following pure density matrices:

\vspace{\baselineskip}
\noindent\resizebox{\linewidth}{!}{%
$\displaystyle
\left\{
	\begin{array}{lll}
		(\gamma^2\text{\textbar}00\rangle+\gamma\lambda\text{\textbar}01\rangle+\gamma\lambda\text{\textbar}10\rangle+\lambda^2\text{\textbar}11\rangle)(\gamma^2\text{\textbar}00\rangle+\gamma\lambda\text{\textbar}01\rangle+\gamma\lambda\text{\textbar}10\rangle+\lambda^2\text{\textbar}11\rangle)^\dagger & & 1-p\\
		(\gamma^2\text{\textbar}00\rangle-\gamma\lambda\text{\textbar}01\rangle+\gamma\lambda\text{\textbar}10\rangle-\lambda^2\text{\textbar}11\rangle)(\gamma^2\text{\textbar}00\rangle-\gamma\lambda\text{\textbar}01\rangle+\gamma\lambda\text{\textbar}10\rangle-\lambda^2\text{\textbar}11\rangle)^\dagger  & & p/3\\
		(\gamma\lambda\text{\textbar}00\rangle+\gamma^2\text{\textbar}01\rangle+\lambda^2\text{\textbar}10\rangle+\gamma\lambda\text{\textbar}11\rangle)(\gamma\lambda\text{\textbar}00\rangle+\gamma^2\text{\textbar}01\rangle+\lambda^2\text{\textbar}10\rangle+\gamma\lambda\text{\textbar}11\rangle)^\dagger  & & p/3\\
		(-\gamma\lambda\text{\textbar}00\rangle+\gamma^2\text{\textbar}01\rangle-\lambda^2\text{\textbar}10\rangle+\gamma\lambda\text{\textbar}11\rangle)(-\gamma\lambda\text{\textbar}00\rangle+\gamma^2\text{\textbar}01\rangle-\lambda^2\text{\textbar}10\rangle+\gamma\lambda\text{\textbar}11\rangle)^\dagger  & & p/3\\
	\end{array} \right.$}
\vspace{\baselineskip}

We note that $X$ and $Y$ coexist in the quantum system, and thus the joint density matrix has been obtained. We already know that $X$ is the cause of $Y$ in this scenario, i.e., $X\to Y$ is the corresponding directed graph. To verify this, we use Algorithm  \ref{alg:qLatentSearch} and \ref{alg:qInferGraph} as we explained earlier in this model.  The results are summarized in Figure \ref{t:DGdepolar}.  $\textcolor{blue}{T}$ means that \qinferg~(Algorithm \ref{alg:qInferGraph}) identifies the direct graph correctly. But, $\textcolor{red}{F}$ means that the algorithm fails to identify the direct graph. In all cases the probability of $X$ be in state $X_1$ is $q=0.4$. 

From a combination of Part I and Part II, we note that for this setup, there are no false positive or false negatives. This shows that the choice of hyperparameters is well suited for the problem, and that the proposed framework is efficient in determining if there is a latent confounder. 

\begin{figure}
    \centering
    \includegraphics{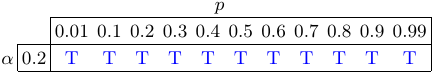}
    \caption{Validation of Direct Graph in Model   \ref{model:depolar} (Part II) with joint density matrix $\rho_{XY}=0.4*\rho_{XY}^{0.6,0.8}+0.6*\rho_{XY}^{1,0}$. }
    \label{t:DGdepolar}
\end{figure}
\end{model}

\section{Why Should We Not Map Quantum to Classical Directly?}\label{sec:mapping}
Here, we show  why classical common entropy approach do not directly apply to the quantum case. 
We emphasize that although a joint density operator (matrix) can be converted to a joint probability distribution (as explained in Example \ref{ex:counterex}), we \textit{lose} some \textit{quantum information} due to the loss of entanglement. We give an example  that shows converting a joint density matrix $\rho_{XY}$ directly to a joint probability distribution $p(X,Y)$, and then applying classical common entropy approach on $p(X,Y)$ will not lead to the correct results.


\begin{algorithm}[!ht]
\caption{Rotational procedure for computing the joint probability distribution of a joint density matrix}\label{alg:rotate}
    \SetAlgoLined
	\footnotesize\KwIn{Joint density matrix of quantum systems $X$ and $Y$ i.e., $\rho_{XY}$.}
	\KwOut{Joint probability distribution $p(X,Y)$ corresponding to the joint density matrix $\rho_{XY}$.}
	\tcc{{Compute eigenvalues and eigenvectors of $\rho_X.$}}
	$[V_1, D_1] = eig(\rho_X)$\;
	\tcc{{Compute eigenvalues and eigenvectors of $\rho_Y.$}}
	$[V_2, D_2] = eig(\rho_Y)$\;
	\tcc{{Rotational procedure}}
	$U= V_1\otimes V_2$\;
	$\rho'_{XY}= U^\dagger \rho_{XY}U$\;
	\textbf{return} $p(X,Y)$ as the entries on the main diagonal of $\rho'_{XY}$.
\end{algorithm}
\begin{example}[Counter Example]\label{ex:counterex} 
    Assume the depolarizing channel as described in Model \ref{model:depolar}, Part II. We already know that $X$ causes $Y$ in this model. 
    To convert the joint density matrix $\rho_{XY}$, we use a rotational procedure explained as follows: Assume that $\rho_{XY}$ is rotated using a unitary matrix $U$. Let us say $\rho_{XY} = U \rho'_{XY} U^\dagger$. So, the joint density matrix $\rho'_{XY}$ is computed as $\rho'_{XY}= U^\dagger \rho_{XY}U$. To compute the unitary matrix $U$ for a given $\rho_{XY}$ we use the eigenspaces of $\rho_X$ and $\rho_Y$, where $\rho_X=\textbf{\textrm{Tr}}_Y(\rho_{XY})$ and $\rho_Y=\textbf{\textrm{Tr}}_X(\rho_{XY})$ are computed by tracing out $Y$ and $X$, respectively. This simple observation enables us to design a procedure that converts a joint density matrix $\rho_{XY}$ to a joint probability distribution $p(X,Y)$ in a way that it takes into account the rotation. This procedure is formally described in Algorithm \ref{alg:rotate}. 
By converting the joint density matrix $\rho_{XY}$ directly to a joint probability distribution $p(X,Y)$, using Algorithm \ref{alg:rotate}, and then applying classical entropic causal inference, i.e., Algorithm \ref{alg:InferGraph} on $p(X,Y)$ we obtain the results represented in Figure \ref{t:mapping} which are opposite to the expected results in all cases. This confirms that classical statistics are not adequate for identification of cause–effect relations in quantum systems due to accessibility of a richer spectrum of causal relations in quantum scenarios.
\begin{figure}
    \centering
    \includegraphics{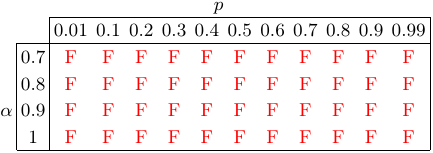}
    \caption{Classical Approach to Identify Direct Graph for Model \ref{model:depolar} does not work.}
    \label{t:mapping}
\end{figure}
\end{example}



\section{Conclusion}\label{sec13}

This paper provides a new approach for quantum entropic causal inference in the presence of hidden common causes. As a part of the approach, an iterative algorithmic solution is provided for the optimization problem that deals with the trade-off between the entropy of the latent quantum system and the quantum conditional mutual information of the observed quantum systems. {We show that the use of quantum density matrix helps achieve significantly better tradeoff even for the classical data.} The approach is validated on quantum noisy links, where the approach detects the expected causal relation or correlation without causation. Our experiments on the synthetic and real classical data confirms that our quantum entropic approach takes advantage of quantum dependency between random variables through density matrices, and as a result it outperforms its  classical counterpart approach. 




\acks{This research was supported by the Defense Advanced Research Projects Agency (DARPA) Quantum Causality [Grant
No. HR00112010008].
}





\clearpage
\bibliography{references}
\appendix
\section{Proof of Theorem \ref{thm:stationary}}\label{sec:appA}
{To prove the theorem, we first write the objective function ($L=I_Q(X;Y\text{\textbar}Z)+\beta S(Z)$) in Equation \eqref{eq:qlossfunc} more explicitly
in terms of the optimization variables $\rho_{Z\text{\textbar}X,Y}$ as follows:
\begin{equation}\label{eq:newloss}
\begin{split}
L &= I_Q(X;Y\text{\textbar}Z)+\beta S(Z)\\
 & = S(XZ) + S(YZ) - S(Z) - S(XYZ) + \beta S(Z)\\
 &=  S(XZ) + S(YZ) - S(XYZ) + (\beta-1) S(Z)\\
 &= S(X) + S(Z\text{\textbar}X)+S(Y)+S(Z\text{\textbar}Y)-S(XY)-S(Z\text{\textbar}X,Y)+(\beta-1) S(Z)\\
 &= S(Z\text{\textbar}X)+S(Z\text{\textbar}Y)-S(Z\text{\textbar}X,Y)+(\beta-1) S(Z)+I_Q(X;Y) 
\end{split}
\end{equation}

To find the stationary points of the loss function $L$, we take its first matrix derivative w.r.t. $\rho_{Z\text{\textbar}X,Y}$ and set it to zero. Let's start with the first term of the new loss function $L$ in Equation \ref{eq:newloss}, i.e., $S(Z\text{\textbar}X)=S(\rho_{Z\text{\textbar}X})$. We have:
\begin{equation}\label{eq:firstterm}
\begin{split}
 \frac{\partial S(\rho_{Z\text{\textbar}X})}{\partial \rho_{Z\text{\textbar}X,Y}} &= \frac{\partial S(\rho_{Z\text{\textbar}X})}{\partial \rho_{Z\text{\textbar}X}}\frac{\partial \rho_{Z\text{\textbar}X}}{\partial \rho_{Z\text{\textbar}X,Y}}\\
 &= (I + \log(\rho_{Z\text{\textbar}X}))\frac{\partial (Tr_Y((\rho^{1/2}_{Y\text{\textbar}X}\otimes I_Z)\rho_{Z\text{\textbar}X,Y}(\rho^{1/2}_{Y\text{\textbar}X}\otimes I_Z)))}{\partial \rho_{Z\text{\textbar}X,Y}}\\
 &= (I + \log(\rho_{Z\text{\textbar}X}))(I)\\
 &= I + \log(\rho_{Z\text{\textbar}X})
 \end{split}
\end{equation}

Note that in Equation \ref{eq:firstterm}, we used matrix calculus as follows: $\frac{\partial \textrm{tr}(AXB)}{\partial X}= BA$, where $A$ and $B$ are not a function of $X$. Also, for a joint probability distribution $p(X,Y)$ we have: $\sum_{y\in Y}p(y\text{\textbar}x)=1$. Similarly, we have the following identity for matrix version of this equation, i.e., $Tr_Y(\rho_{Y\text{\textbar}X})=I$. Following  similar matrix calculations for other terms in Equation \ref{eq:newloss}, we obtain:
\begin{equation}\label{eq:partialD}
\begin{split}
 \frac{\partial L}{\partial \rho_{Z\text{\textbar}X,Y}} &= [I + \log(\rho_{Z\text{\textbar}X})]+[I + \log(\rho_{Z\text{\textbar}Y})]-[I + \log(\rho_{Z\text{\textbar}X,Y})]+(\beta-1)[I + \log(\rho_{Z})]
 \end{split}
\end{equation}

By solving $\frac{\partial L}{\partial \rho_{Z\text{\textbar}X,Y}}=0$ from Equation \ref{eq:partialD}, assuming that all density matrices are  positive definite\footnote{Even though the assumption of positive definiteness may not always be valid, we can replace $\rho$ with  $(1-\epsilon)\rho + \epsilon I$ for very small $\epsilon$ to alleviate the issue in the approach and the algorithm. This will allow for the existence of the logarithm of the matrices. }, we obtain:
$$\rho_{Z\text{\textbar}X,Y}= \exp(\log(\rho_{Z\text{\textbar}X})+\log(\rho_{Z\text{\textbar}Y})+(\beta-1)\log(\rho_{Z}))$$
This means a point is a stationary point of the loss function $L$ if and
only if it is a stationary point of \qlsearch~(Algorithm \ref{alg:qLatentSearch}).}

\end{document}